%% file: main.tex
\documentclass[a4paper,fleqn]{cas-sc}
\usepackage[square,numbers]{natbib}

\usepackage[switch,mathlines]{lineno}

\newif\ifarxiv
\arxivtrue

\newif\ifdoubleblind

\usepackage{amssymb}
\usepackage[utf8]{inputenc}
\UseRawInputEncoding

\input{commands/cas-abbrev-fix}
\input{commands/cas-prelim-fix}
\input{commands/citeabbrev}

\usepackage{lipsum}

\usepackage{tabularx}
\usepackage{xltabular}
\usepackage{booktabs}
\usepackage{pdflscape}
\usepackage{rotating}
\usepackage{longtable}
\usepackage{multirow}
\usepackage{caption}
\usepackage{float}
\usepackage{ragged2e}
\usepackage{caption}
\usepackage{cleveref}
\usepackage{subfig}
\usepackage{siunitx}
\DeclareSIUnit \decibelA {dB(A)}
\DeclareSIUnit \decibelC {dB(C)}
\DeclareSIUnit \soneGF {soneGF}
\DeclareSIUnit \acum {acum}
\DeclareSIUnit \asper {asper}
\DeclareSIUnit \vacil {vacil}
\DeclareSIUnit \tuhms {tuHMS}

\usepackage{xcolor}
\definecolor{set1_1}{RGB}{228,26,28}
\definecolor{set1_2}{RGB}{55,126,184}
\definecolor{set1_3}{RGB}{77,175,74}
\definecolor{den_1}{RGB}{239,209,0}
\definecolor{den_2}{RGB}{78,184,123}
\definecolor{den_3}{RGB}{0,127,196}
\definecolor{loaclr}{RGB}{152, 78, 163}
\definecolor{mclr}{RGB}{255, 127, 0}

\usepackage{hyperref}
\hypersetup{
    colorlinks=true,
    linkcolor=blue,
    filecolor=magenta,      
    urlcolor=cyan,
    pdftitle={Overleaf Example},
    pdfpagemode=FullScreen,
    }
\usepackage{comment}
\usepackage{setspace}
\usepackage[section]{placeins}
\usepackage{float}

\newif\ifshowchanges

\definecolor{lightgray}{RGB}{200,200,200}
\ifshowchanges
    \usepackage[commandnameprefix=always]{changes}
    \setcommentmarkup{\marginpar{\small\centering\color{purple}\textbf{[{#1}]}}}
    \setdeletedmarkup{ {\color{lightgray}\sout{#1}}}
\else
    \usepackage[final,commandnameprefix=always]{changes}
\fi

\input{commands/shorthand}

\begin{document} 
\let\WriteBookmarks\relax
\def\floatpagepagefraction{1}
\def\textpagefraction{.001}

\input{body/00frontmatter}


\doublespacing

\input{body/01introduction}

\input{body/02method}

\input{body/03results}

\input{body/04discussion}

\input{body/05conclusion}

\input{body/98endmatter}

\input{body/99appendix.tex}

\twocolumn

\bibliographystyle{model1a-num-names.bst}
\bibliography{references-zot}





\end{document}

%% file: commands/cas-abbrev-fix.tex
\ExplSyntaxOn
\RenewDocumentCommand \eadauthor {} 
    { 
      \seq_map_inline:Nn \l_stm_au_seq 
        { 
            \regex_extract_once:nnNTF {(\w)\w*-(\w)} { ##1 } \l_stm_au_fn_seq
            { 
                \seq_pop_left:NN \l_stm_au_fn_seq \temp_var
                \seq_use:Nn \l_stm_au_fn_seq { .- }
                { . } 
            }
            { 
                \regex_match:nnTF { \. } { ##1 } 
                { ##1 }
                { \tl_head:n {##1}. }
            }
      }{ ~\l_stm_au_sn_seq }
    }
\ExplSyntaxOff

%% file: commands/cas-prelim-fix.tex
\ExplSyntaxOn
\makeatletter

\RenewDocumentEnvironment { PrelimsAbstract } { O{} }
  {\parindent=0pt 
   { \fontsize{14pt}{16pt}\selectfont #1 }\par
   \vskip12pt
   { \fontsize{12pt}{14pt}\bfseries\selectfont\casprelimstitle } \par
   \vskip6pt
   \ifnum\theblind>0\relax
     \vspace*{\the\baselineskip}
   \else  
     \seq_use:Nn \g_stm_prelimsau_seq { ,\ }
   \fi  
   \vskip12pt
   \par 
  } 
  {}
  
\makeatother
\ExplSyntaxOff

%% file: commands/citeabbrev.tex
\newif\ifabbreviation
\pretocmd{\thebibliography}{\abbreviationfalse}{}{}
\AtBeginDocument{\abbreviationtrue}
\DeclareRobustCommand\acroauthor[2]{%
  \ifabbreviation
    \ifcsname acroused@#2\endcsname
      #2%
    \else
      #1%
      ~[\mbox{#2}]
      \expandafter\gdef\csname acroused@#2\endcsname{}%
    \fi
  \else
    \ifcsname bibacroused@#2\endcsname
        \mbox{#2}%
    \else
        \mbox{#1}~(\mbox{#2})%
        \expandafter\gdef\csname bibacroused@#2\endcsname{}%
    \fi
  \fi
}

%% file: commands/shorthand.tex
\DeclareSIUnit \decibelA {dB(A)}
\DeclareSIUnit \decibelC {dB(C)}
\DeclareSIUnit \soneGF {soneGF}
\DeclareSIUnit \acum {acum}
\DeclareSIUnit \asper {asper}
\DeclareSIUnit \vacil {vacil}
\DeclareSIUnit \tuhms {tuHMS}

\newcommand{\db}[1]{\ifdoubleblind \{\texttt{masked}\}\else #1\fi}

\newcommand{\gfp}[0]{\texttt{GND}}
\newcommand{\rtgp}[0]{\texttt{ROOF}}

\newcommand{\amb}[0]{\texttt{AMB}}
\newcommand{\amss}[0]{\texttt{AMSS}}
\newcommand{\rtrmanova}[0]{2ME-RT-RMANOVA}
\newcommand{\rmanova}[0]{2ME-RMANOVA}

\newcommand{\fas}[0]{\textit{PRSS}\textsubscript{Fas}}
\newcommand{\com}[0]{\textit{PRSS}\textsubscript{Com}}
\newcommand{\ba}[0]{\textit{PRSS}\textsubscript{BA}}
\newcommand{\ec}[0]{\textit{PRSS}\textsubscript{EC}}
\newcommand{\es}[0]{\textit{PRSS}\textsubscript{ES}}

\newcommand{\nat}[0]{\textit{DOM}\textsubscript{Nat}}
\newcommand{\hum}[0]{\textit{DOM}\textsubscript{Hum}}
\newcommand{\noi}[0]{\textit{DOM}\textsubscript{Noi}}

\newcommand{\pa}[0]{\textit{PA}}
\newcommand{\na}[0]{\textit{NA}}
\newcommand{\osq}[0]{\textit{OSQ}}
\newcommand{\appr}[0]{\textit{APPR}}
\newcommand{\pln}[0]{\textit{PLN}}
\newcommand{\isopl}[0]{\textit{ISOPL}}
\newcommand{\isoev}[0]{\textit{ISOEV}}

\newcommand{\laeq}[0]{$L_\text{A,eq}$}
\newcommand{\lceq}[0]{$L_\text{C,eq}$}
\newcommand{\nnf}[0]{$N_\text{95}$}
\newcommand{\dba}[1]{\SI{#1}{\decibelA}}
\newcommand{\dbc}[1]{\SI{#1}{\decibelC}}

%% file: body/00frontmatter.tex
\newcommand*{\papertitle}{Automating Urban Soundscape Enhancements with AI: In-situ Assessment of Quality and Restorativeness in Traffic-Exposed Residential Areas}
\shorttitle{\papertitle}    
\shortauthors{Lam et al.}  

\title[mode=title]{\papertitle}

\author[eee]{Bhan Lam}[
    auid=000,
    bioid=001,
    degree=PhD,
    orcid=0000-0001-5193-6560]
\ead{blam002@e.ntu.edu.sg}
\corref{c}\cortext[c]{Corresponding author}
\credit{Conceptualization, Methodology, Software, Validation, Formal analysis, Investigation, Project administration, Data Curation, Writing - Original Draft, Writing - Review \& Editing, Visualization, Supervision}

\author[eee]{Zhen-Ting Ong}[orcid=0000-0002-1249-4760]
\credit{Project administration, Investigation, Data Curation}

\author[eee]{Kenneth Ooi}[orcid=0000-0001-5629-6275]
\credit{Methodology, Investigation, Software, Data Curation, Validation, Formal analysis, Visualization, Writing - Original Draft, Writing - Review \& Editing}

\author[eee]{Wen-Hui Ong}
\credit{Software, Data Curation}

\author[eee]{Trevor Wong}
\credit{Software, Data Curation}

\author[gt,eee]{Karn~N. Watcharasupat}[orcid=0000-0002-3878-5048]
\credit{Software, Writing - Review \& Editing}

\author[hdb]{Vanessa~Boey}
\credit{Project administration, Investigation, Data Curation, Resources}

\author[hdb]{Irene~Lee}[degree=PhD]
\credit{Conceptualization, Writing - Review \& Editing, Supervision, Resources, Funding acquisition}

\author[cnu]{Joo Young Hong}[orcid=0000-0002-0109-5975, degree=Ph.D.]
\credit{Methodology, Writing - Review \& Editing, Supervision}

\author[ucl]{Jian Kang}[orcid=0000-0001-8995-5636, degree=Ph.D.]
\credit{Methodology, Writing - Review \& Editing, Supervision}

\author[nbs,hku]{Kar Fye Alvin Lee}[degree=PhD,orcid=0000-0003-3774-6714]
\credit{Formal analysis, Writing - Review \& Editing}

\author[nbs]{Georgios Christopoulos}[degree=PhD,orcid=0000-0003-2492-653X]
\credit{Formal analysis, Writing - Review \& Editing, Funding acquisition}

\author[eee]{Woon-Seng Gan}[degree=PhD,orcid=0000-0002-7143-1823]
\credit{Conceptualization, Writing - Review \& Editing, Supervision, Resources, Funding acquisition}

\affiliation[eee]{organization={School of Electrical and Electronic Engineering, Nanyang Technological University},
            addressline={50 Nanyang Avenue}, 
            city={Singapore 639798},
            country={Singapore}}

\affiliation[gt]{
    organization={
        Center for Music Technology,
        Georgia Institute of Technology%
    },
    addressline={J. Allen Couch Building, 840 McMillan St NW}, 
    city={Atlanta},
    postcode={30332}, 
    state={GA},
    country={USA}
}

\affiliation[hdb]{organization={Building \& Research Institute, Housing \& Development Board},
            city={Singapore 738973},
            country={Singapore}}

\affiliation[cnu]{
    organization={
        Department of Architectural Engineering,
        Chungnam National University%
    },
    addressline={34134}, 
    city={Daejeon},
    country={Republic of Korea}
}

\affiliation[ucl]{
    organization={
        UCL Institute for Environmental Design and Engineering, The Bartlett, University College London, Central House,%
    },
    addressline={14 Upper Woburn Place}, 
    city={London WC1H 0NN},
    country={United Kingdom}
}

\affiliation[nbs]{organization={Nanyang Business School, Nanyang Technological University},
            addressline={50 Nanyang Avenue}, 
            city={Singapore 639798},
            country={Singapore}}

\affiliation[hku]{organization={Laboratory of Neuropsychology and Human Neuroscience, Department of Psychology, The University of Hong Kong},
            addressline={Pokfulam Road}, 
            country={Hong Kong}}

\tnotemark[1]
\tnotetext[1]{The research protocols used in this research were approved by the institutional review board of Nanyang Technological University (NTU), Singapore [IRB-2023-399].}

\begin{abstract}
Formalized in ISO 12913, the ``soundscape'' approach is a paradigmatic shift towards perception-based urban sound management, aiming to alleviate the substantial socioeconomic costs of noise pollution to advance the United Nations Sustainable Development Goals. Focusing on traffic-exposed outdoor residential sites, we implemented an automatic masker selection system (AMSS) utilizing natural sounds to mask (or augment) traffic soundscapes. We employed a pre-trained AI model to automatically select the optimal masker and adjust its playback level, adapting to changes over time in the ambient environment to maximize ``Pleasantness'', a perceptual dimension of soundscape quality in ISO 12913. Our validation study involving ($N=68$) residents revealed a significant \SI{14.6}{\percent} enhancement in ``Pleasantness'' after intervention, correlating with increased restorativeness and positive affect. Perceptual enhancements at the traffic-exposed site matched those at a quieter control site with \SI{6}{\decibelA} lower $L_\text{A,eq}$ and road traffic noise dominance, affirming the efficacy of AMSS as a soundscape intervention, while streamlining the labour-intensive assessment of ``Pleasantness'' with probabilistic AI prediction.

\end{abstract}


\maketitle

%% file: body/01introduction.tex
\section{Introduction}
\label{sec:Introduction}

\subsection{Background and motivation}
\label{sec:Introduction/Background and motivation}

In urban environments, road traffic noise poses significant annual economic burdens, rivaling those of road accidents, as evidenced by estimates in England (\textsterling 7 billion) and across Europe (\texteuro 38 billion) \citep{WHO2018,defra2014,King2022a}. Beyond economic concerns, the documented adverse physical and mental health effects of urban noise warrant urgent mitigation \citep{WHO2018,Fink2019,newbury_air_2024,hahad_noise_2024}. For instance, even a modest reduction of \dba{5} in noise levels has been projected to yield substantial annual economic benefits from adverse health effects in the United States, totaling \$3.9 billion \citep{King2022a}.

Crucially, mere reductions in sound pressure levels (SPLs) may not uniformly translate into perceptual improvements. Considerable variations in annoyance and comfort levels have been found among individuals exposed to identical SPLs, highlighting the complexity of the urban ``soundscape'' perception \cite{Guski2017,kang_urban_2007,Kang2017,Kang2016}.

The soundscape approach, formalized in the ISO 12913 series  \citep{iso12913-1, iso12913-2, iso12913-3}, offers a holistic strategy for urban sound management, aligning with the United Nations Sustainable Development Goals (SDGs), particularly SDG 3 (well-being) and SDG 11 (sustainable cities), by accounting for how humans perceive and experience their aural environments, in context. The significance of this approach is echoed by the United Nations Environment Program Frontiers 2022 report, which emphasized the need to mitigate unwanted noise while harnessing the health-promoting benefits of natural sounds \citep{Buxton2021,UNEP2022,Fisher2023}.

\subsection{Soundscape augmentation for road traffic noise}

Soundscape augmentation emerges as a viable intervention technique under the ISO 12913 paradigm. Additional sounds, known as ``maskers'', are augmented to existing soundscapes through loudspeakers or electroacoustic systems. In prior art, maskers used in traffic-exposed urban areas typically comprise natural sounds, such as wind sounds \citep{Jeon2010}, sounds from animals (such as birds \citep{Hao2016,Hong2020b} and insects \citep{Hong2017ICSV}), water sounds (such as man-made water features \citep{Yang2019}, natural waterfalls \citep{Jeon2012}, waves \citep{Radsten-Ekman2013a}, and streams \citep{You2010,Lam2023}), and corresponding mixtures \citep{cerwen_urban_2016}.

Specifically, \citet{Calarco2024} modeled the propagation of water feature sounds in a park exposed to traffic noise, defining optimal listening zones where water sounds were not less than \SI{3}{\decibel} below the traffic noise levels \citep{Jeon2012}. They found that the optimal zone decreases with increasing traffic noise levels, in addition to variations in preference among various water feature varieties. Conversely, a laboratory study by \citet{Nilsson2010} found a significant reduction in traffic noise perception only when the fountain sound exceeded road noise by at least \SI{10}{\decibel}. A \SI{9}{\percent} improvement in overall sound quality post-augmentation was reported, favoring compositions with songbirds at varying volumes. Furthermore, \cite{Chau2023} found that participants were more likely to be highly annoyed when traffic noise was perceived to be the dominant sound source under augmentation with birdsongs and stream sounds. On the contrary, a separate virtual reality (VR)-study found no evidence that any particular birdsong composition augmented to soundscapes of a Swedish park reduced stress levels \citep{Hedblom2019a}. \citet{VanRenterghem2020} explored real-world soundscape augmentation in a traffic-exposed park by inviting participants to customize natural sound samples emitted from a hidden speaker to their preference. Hence, it would be naive to assume that every bird masker (or every masker from the same class in general) would improve the quality of a given soundscape, thereby necessitating some form of selection process to effect a desired perceptual change.

Moreover, few studies have extended their findings into soundscape augmentation systems for road traffic noise in real-life urban environments. Installing and uninstalling speakers in a soundscape augmentation system can also be more cost-effective and conducive to the surrounding environment as compared to alternative methods of noise mitigation such as noise barriers, which require physical space and may be more difficult to retrofit to existing urban areas \citep{Lam2021}.

\subsection{Masker selection methods for soundscape augmentation}


One real-life soundscape augmentation system was explored by \citet{VanRenterghem2020} in a park in Ghent, Belgium, where road traffic noise was dominant. Participants composed their own maskers by adjusting the playback levels of eight natural sound samples emanating from a hidden loudspeaker, then evaluated both the original and augmented soundscapes. The study observed a mean improvement of 0.36 unit (\SI{9}{\percent}) in overall sound quality on a 5-point scale, with most participants preferring the sounds of house sparrows and mixed songbirds.

Similar effects may also be observed even if the loudspeaker or speaker systems are visible to the participants. \citet{Hong2021b} conducted a study with participants standing at pedestrian walkways near roads, adjusting the soundscape-to-masker (SMR) ratio of birdsong and fountain recordings reproduced by down-firing speakers of a mixed-reality device. The recordings were accompanied either by a hologram matching their source (a bird for the birdsong and a jet-and-basin fountain for the fountain) or by a visible speaker. Participants adjusted the SMR to a level they found most preferable for masking traffic noise. The study found no significant differences in the chosen SMRs or the resultant ratings of overall soundscape quality and perceived loudness of traffic noise between the hologram and speaker conditions. In addition,  \citet{Regazzi2021} used the frequency spectrum of transformer noise in a residential area to create a natural sound masker, aiming to equalize tonal frequencies when reproduced over speakers. This demonstrates the effectiveness of speakers in soundscape augmentation, despite the potential lack of realism compared to real-life sources.

However, these methods require participant involvement or expert input to generate optimal maskers and playback gains, which may not be practical for long-term deployments. Changing soundscape characteristics over time can render previously optimal maskers suboptimal.

Alternatively, model-based approaches offer the potential for generalizability across scenarios. For instance, \citet{Lenne2020} optimized masker playback locations indoors based on room acoustics simulations, while others have incorporated physical models for real-time augmentation of footstep sounds in virtual-reality soundscapes \citep{Nordahl2010, Nordahl2011, Turchet2013}. \citet{Suhanek2019} optimized the ``total distraction coefficient'' to select appropriate songs as maskers for park and expressway soundscapes, but only theoretically validated their masker choices. Despite the promise, model-based approaches remain sparse in the literature, particularly in the context of road traffic noise, and none have been developed using the ISO 12913 framework. 

Automated masker selection methods could enhance efficiency by reducing the time and labor involved in human evaluation, while also adapting to changing soundscapes. The success of automated masker selection relies on the availability of reliable models to predict affective responses, such as ``\textit{Pleasantness}'' (\isopl) \citep{iso12913-3} or restorativeness \citep{Payne2013,Payne2018}, which are crucial for enhancing acoustic comfort. To date, few prediction models for multidimensional indicators such as \isopl\ have been developed \citep{Lionello2020,Hou2023a,Watcharasupat2022,Ooi2022,Ooi2023b}, and interventions based on enhancing \isopl\ are lacking \citep{Moshona2024}.

\subsection{Research questions}\label{sec:research_questions}
Addressing these gaps, we utilize our probabilistic \isopl\ prediction model, trained on our large-scale dataset of perceptual responses to soundscapes \citep{Ooi2023a}, to deploy and validate a proof-of-concept model-based automatic masker selection system (AMSS) at a traffic-exposed residential site. Operating autonomously, the AMSS augments the soundscape to maximize \isopl. Through in-situ validation, we aim to assess the impact of AMSS on soundscape quality, its influence on related perceptual dimensions, and its correlation with objective acoustic metrics. Specifically, we seek to answer the following research questions:

\begin{enumerate}
    \item[RQ1.] To what extent can the soundscape quality of a traffic-exposed site be modified by the AMSS?
    
    
    
    \item[RQ2.] What impact does optimizing a soundscape intervention to improve \isopl\ have on other soundscape-related perceptual dimensions, such as restorativeness, perceived loudness, and \isoev? 
    
    
    
    \item[RQ3.] How do perceptual changes induced by the AMSS correlate to objective (psycho)acoustic metrics? 
    
    
\end{enumerate}

\label{sec:Introduction/Research questions}

%% file: body/02method.tex
\section{Method}
\label{sec:Method}

The in-situ validation study was conducted between 1 August 2023 and 30 November 2023, and prior to participant recruitment and experimentation, formal ethical approval was obtained from the Institutional Review Board at \db{Nanyang Technological University} (Reference number IRB \db{2023-399}). The study administrators strictly adhered to the approved methodology, and informed consent was obtained from all participants prior to the start of the experiment.

\subsection{Study sites}\label{sec:study_sites} 

\begin{figure}
    \begin{minipage}{.49\textwidth}
        \centering
        \subfloat[]{\label{fig:gnd}\includegraphics[width=0.95\columnwidth]{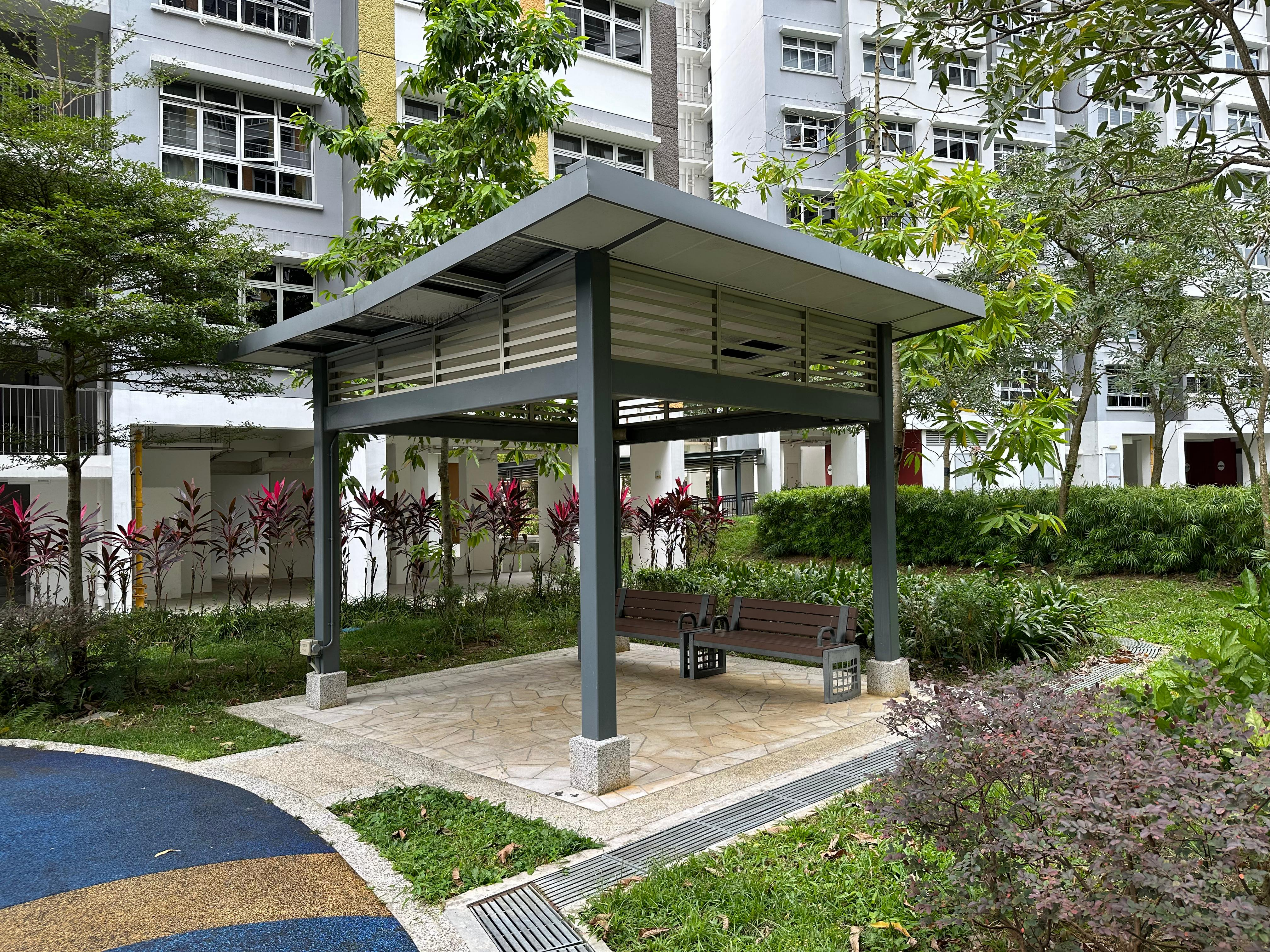}}
    \end{minipage}
    \begin{minipage}{.49\textwidth}
        \centering
        \subfloat[]{\label{fig:roof}\includegraphics[width=0.95\columnwidth]{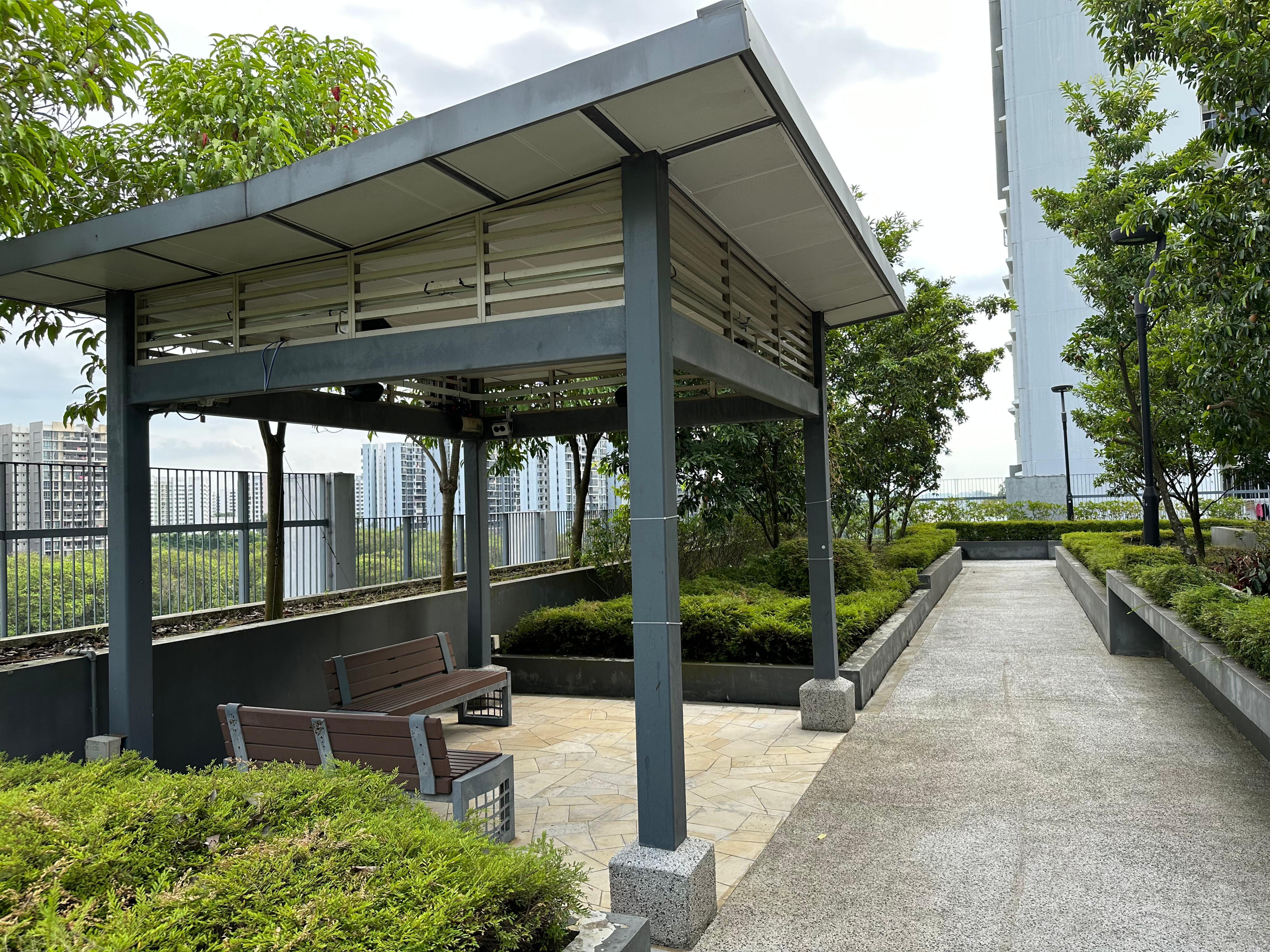}}
    \end{minipage} \par
    \centering
    \subfloat[]{\label{fig:amssdiagram}\includegraphics[width=1\columnwidth]{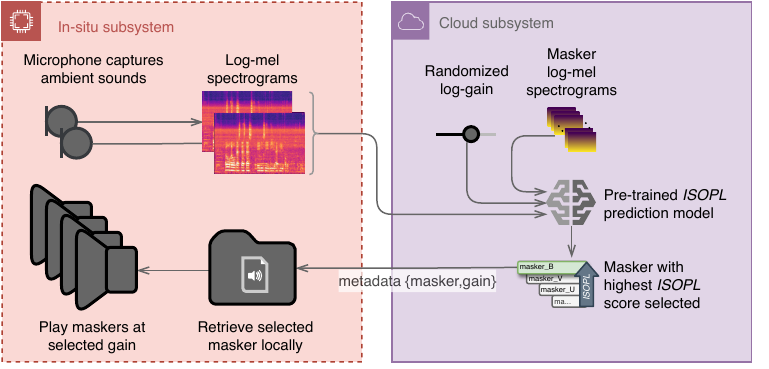}}

    \caption{Study sites in a public residential estate in Singapore: (a) A ground-floor pavilion (\gfp) in the outdoor recreational area at coordinates \db{(1.401358, 103.895427)}. (b) A rooftop garden pavilion (\rtgp) situated atop an 8-storey multi-storey car park at GPS coordinates \db{(1.343373, 103.686134)}. (c) An overview of the end-to-end process of the automatic masker selection system (AMSS)}
    \label{fig:pcp-pictures}
\end{figure}

The study sites were two distinct pavilions within a public residential estate in \db{Singapore}, as shown in \Cref{fig:pcp-pictures}. Both pavilions were identical in design, but were situated at different locations in the estate. 

The first study site was a ground-floor (``\gfp'') pavilion positioned at street level adjacent to a children's playground and fitness area. The \gfp\ was situated amidst six residential apartment blocks, which were in turn surrounded by and served as a physical barrier to a minor 2-lane road (\SI{60}{\meter} away from the pavilion) with light traffic. As a control site, no AMSS was deployed at the \gfp.

The second study site was a rooftop (``\rtgp'') garden pavilion positioned near the periphery of rooftop garden atop an 8-storey multi-storey car park (MSCP), which bordered a major 8-lane expressway with heavy traffic. The \rtgp\ was positioned \SI{30}{\meter} above street level and was flanked by a 2-lane slip road (\SI{50}{\meter} away) leading out from a major 6-lane expressway (\SI{70}{\meter} away). The AMSS was physically deployed at the \rtgp, with four loudspeakers (Moukey M20-2, DONNER LLC, FL, USA) affixed to the pavilion roof (at a height of \SI{2.5}{\meter} above the ground of the pavilion) in a square of length \SI{2.2}{\meter} for the playback of maskers, which were automatically selected and reproduced according to the method described in \Cref{sec:automatic_masker_selection}. A customized Internet-of-Things (IoT)-based infrastructure was used for the deployed AMSS, as detailed by \cite{Wong2022}. The placement of the hardware of AMSS did not physically or visually block any ingress or egress routes to the \rtgp.

\subsection{Design of in-situ validation experiment}\label{sec:experimental_design}

\begin{figure}
    \centering
    \includegraphics[width=\columnwidth]{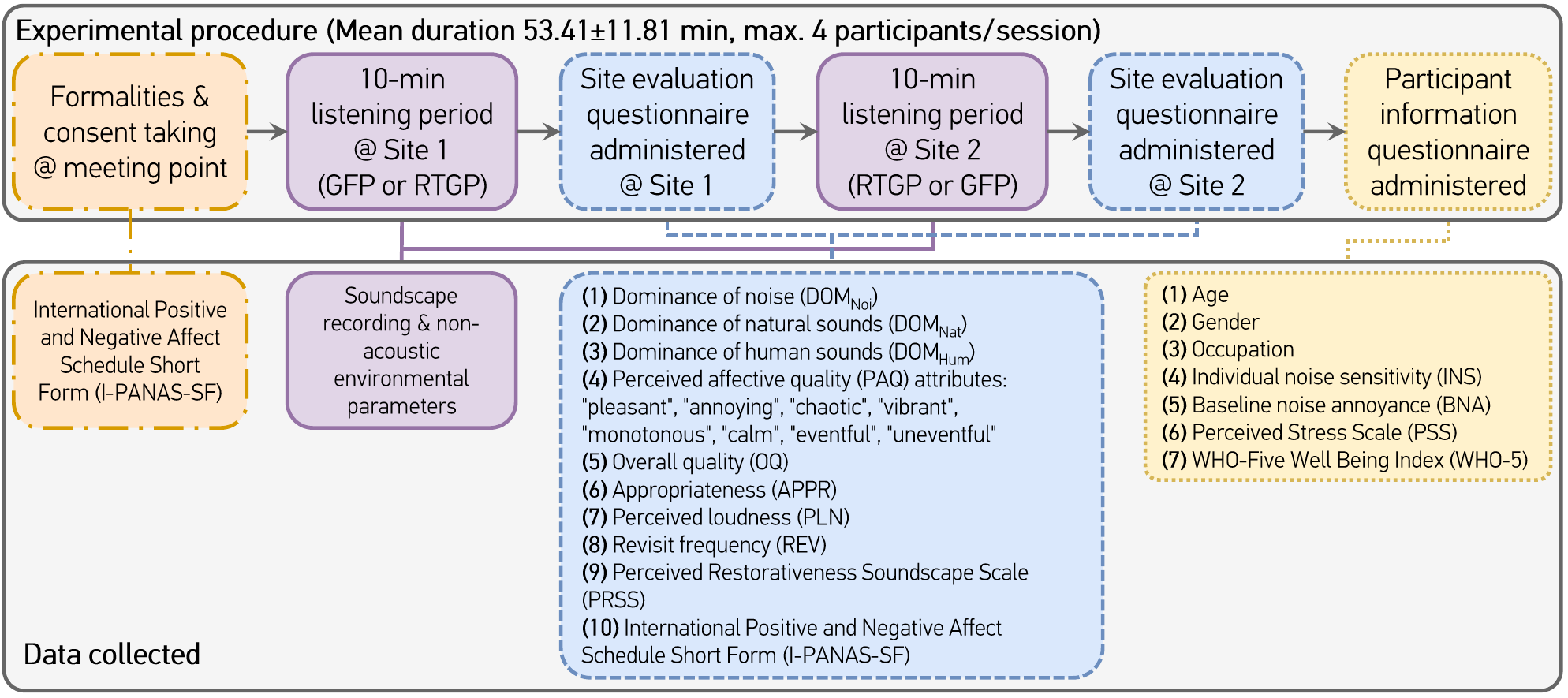}
    \caption{Overview of experimental procedure and data collected from participants for the in-situ validation experiment.}
    \label{fig:experimental-procedure}
\end{figure}

To investigate the influence of the AMSS on soundscape perception, we employed a within-between design. Participants were allocated randomly into two independent groups (between factor): the ``AMSS'' (\amss) and the ``Ambient'' (\amb) group. In both groups, participants evaluated the soundscapes at both the \gfp\ and \rtgp\ (within factor) in a randomized order. However, the AMSS was turned on (i.e., soundscape augmentation was performed according to the method described in \Cref{sec:automatic_masker_selection}) for the \amss\ group at the \chreplaced[comment=R1.2 R2.4]{\rtgp}{RTGP} and turned off (i.e., no soundscape augmentation was performed) for the \amb\ group at the \rtgp. As explained in \Cref{sec:study_sites}, the AMSS was not deployed at the \gfp, so the evaluations at the \gfp\ for participants in both the \amss\ and \amb\ groups corresponded to that of the ambient environment at the \gfp. Given the communal nature of the public space, each session accommodated up to four participants, aligning with the maximum seating capacity of the pavilions. On average, there were 1.53 $\pm$ 0.80 participants per session, and an overview of the experimental procedure for each session is illustrated in \Cref{fig:experimental-procedure}.

At the onset of each session, participants convened at the meeting point (MP) for the requisite consent process, a briefing on the study protocol, and hands-on training with the electronic form used for the evaluations. The International Positive and Negative Affect Schedule Short Form (I-PANAS-SF) \citep{Thompson2007a} was also administered at the meeting point. To prevent undue bias in evaluation, participants were not informed whether they had been placed into the \amss\ or \amb\ group, and were also not informed about the presence of the AMSS system at the \rtgp. Within each study site, participants were initially directed to listen to the pavilion's soundscape for \SI{10}{\minute} without engaging in any other activities and without interacting with each other. Subsequently, they used their personal mobile devices to complete an electronic evaluation form. To ensure clarity, study administrators reiterated the following instructions verbatim to participants before the listening period:

\begin{quote}
    \textit{We will be assessing the sound environment within the pavilion. Over the next 10 minutes, immerse yourself in the surrounding sounds. Choose to sit or stand, but minimize movements to avoid disturbing others. Refrain from using your phone or engaging in other activities. Focus on the types of sounds and your emotional responses, considering the pavilion's context for rest and relaxation.}
\end{quote}

During the \SI{10}{\minute} listening period, the acoustic environment experienced by the participants was captured using a binaural microphone (TYPE 4101-B, Hottinger Brüel \& Kjær A/S, Virum, Denmark) equipped with a windscreen. This microphone was coupled with a data recorder (SQobold, HEAD acoustics GmbH, Herzogenrath, Germany). Ensuring data precision and uniformity, the binaural recording equipment underwent calibration using an IEC 60942 class 1 calibrator (42AG, G.R.A.S. Sound \& Vibration A/S, Holte, Denmark). The responsibility of wearing and operating this equipment rested with a single experiment administrator during each session, with a total of four unique administrators overseeing the entire 4-month study duration.

In alignment with ISO 12913-2 \citep{iso12913-2}, environmental data was systematically collected during the \SI{10}{\minute} listening period. Temperature and humidity readings were obtained from a combined digital humidity and temperature sensor (BME280, Bosch Sensortec GmbH, Reutlingen, Germany), while luminance data was captured by an optical sensor (LTR-559ALS-01, LITE-ON Technology Corp., Taiwan) integrated into the AMSS system at the \rtgp, all at \SI{10}{\minute} intervals. Additionally, wind speed, 24-\si{\hour} pollutant standards index (PSI), and PM2.5 readings were sourced from the nearest weather station via the \db{Singapore} Meteorological Service, also recorded at \SI{10}{\minute} intervals. Detailed specifications regarding the metrics, range, accuracy, and resolution of the measurement instruments are delineated in \Cref{tab:instrument}. 

\input{tables/instruments}

After the \SI{10}{\minute} listening period, the participants were instructed to complete the I-PANAS-SF questionnaire. Thereafter, participants received the following instruction:

\begin{quote}
    \textit{This evaluation is about the surrounding sound environment you just experienced in the past 10 minutes. Answer the following questions by recalling the sounds you experienced in the 10 minutes.}
\end{quote}

The questions formed the site evaluation questionnaire, which prompted participants to rate (1) the dominance of noise (\noi), (2) the dominance of natural sounds (\nat), (3) the dominance of human sounds (\hum), (4) the 8 attributes corresponding to perceived affective quality (PAQ) in the Method A questionnaire of ISO 12913-2, (5) the overall soundscape quality (\osq), (6) appropriateness (\appr), and (7) perceived loudness (\pln) on 5-point scales, on top of the items in the 18-item Perceived Restorativeness Soundscape Scale by \cite{Payne2018}, which were on 7-point scales. The PRSS consists of four main dimensions: \textit{Fascination} (\fas), \textit{Being-Away} (\ba), \textit{Compatibility} (\com), and \textit{Extent}, which consists of two sub-dimensions: \textit{Extent-Coherence} (\ec) and \textit{Extent-Scope} (\es). The precise wording of each item in the site evaluation questionnaire is provided in \ref{sec:questions}, \Cref{tab:stimuliquestion}.

Considering the fatigue and relevance of terms within the local context, the 18-item PRSS scale utilized in this study underwent modification by omitting or consolidating 7 of the 23 items from the PRSS scale with specific framing outlined by \citet{Payne2018}. These adjustments are detailed in \Cref{tab:prss}.

At the end of the soundscape evaluation at the second site, participants completed an additional participant information questionnaire covering basic demographics (gender, age, occupation) and self-reported assessments on the (1) individual noise sensitivity (INS) \citep{weinstein_individual_1978}, (2) baseline noise annoyance (BNA) \citep{iso156662021}, (3) Perceived Stress Scale (PSS-10) \citep{Cohen1983}, and (4) WHO-Five Well-being Index (WHO-5) \citep{WorldHealthOrganization1998}. exact wording of every item in the participant information questionnaire can be found in \Cref{tab:participantquestion}. The experimental procedure averaged 53.41 $\pm$ \SI{11.81}{\minute} to complete. 

\subsection{Stimuli and automatic masker selection}\label{sec:automatic_masker_selection}

As explained in \Cref{sec:experimental_design}, only the \amss\ group experienced augmented soundscapes with maskers presented over four loudspeakers in the \rtgp. The maskers were selected from the bank of maskers in the ARAUS dataset \citep{Ooi2023a}, comprising 280 different processed recordings of birds, water, wind, traffic, and construction as \SI{30}{\second} mono tracks.

Specifically, a pre-trained artificial intelligence (AI) model decoupling the spectrograms of the existing soundscape, masker, and playback gain \citep{Watcharasupat2022} was used in the AMSS to pick maskers and corresponding gain values, in intervals of \SI{30}{\second}. The model was trained on the \num[group-separator = {,}]{25440} subjective responses to augmented urban soundscapes in the ARAUS dataset to predict distributions of ISO Pleasantness (\isopl), as defined in ISO 12913-3 \citep{iso12913-3}, which the AMSS then used to select a masker-gain combination at each interval to maximise the \isopl\ of the existing soundscape at the \rtgp. An overview of the AMSS system is depicted in \Cref{fig:amssdiagram}. 

\chadded[comment=R1.1]{The model training and validation data consist of participant evaluations of 42 \SI{30}{\second} excerpts, randomly selected from 234 ``base soundscapes'' in the Urban Soundscapes of the World dataset, and/or augmented with 280 curated maskers from the \texttt{Freesound} and \texttt{xeno-canto} databases. A five-fold cross-validation approach was used to train the probabilistic perceptual attribute predictor (PPAP) model \citep{Watcharasupat2022, Ooi2022}, with each fold carefully designed to include tracks with similar psychoacoustic distributions to minimize bias, as described in \citep{Ooi2023a}. Model evaluation on an independently recorded test set demonstrated that the PPAP and a CNN-based deep learning approach for predicting \isopl\ significantly outperformed a linear elastic net model using psychoacoustic parameter inputs \citep{Ooi2023a}. The trained PPAP model is deployed as an inference node in the cloud, where it serves as the \isopl\ predictor for the AMSS. } 

\chadded[comment=R2.2]{Since the PPAP was trained with \SI{30}{\second} log-mel spectrogram inputs of soundscapes and maskers, the AMSS maskers were updated in \SI{30}{\second} intervals, totalling 20 masker-gain combinations in the \SI{10}{\minute} listening period, where the optimal maskers were selected based on the soundscape of the previous \SI{30}{\second}.} At each \SI{30}{\second} interval, the AMSS randomly picked 5 gain values from a log-normal distribution for each masker candidate in the masker bank, with the log-gains being normally distributed with mean $-2.0$ and standard deviation $1.5$. These values match the distribution of log-gains in the ARAUS dataset maskers when calibrated to an SPL of \SI{65}{\decibelA} and correspond to five possible SMRs when applied to the maskers upon playback. For each of these masker-gain combinations, the AI model gave as initial output the predicted \isopl\ distributions as though they were used to augment the existing soundscape. Then, the masker-gain combinations were ranked in terms of the predicted improvement in \isopl\ via the estimation scheme described by \cite{Ooi2022}. Lastly, the top-ranked masker-gain configuration was reproduced across the four loudspeakers in the deployed AMSS with each loudspeaker playing back the same masker at the same SPL corresponding to the gain value.

The sound level output of each loudspeaker was previously calibrated for each masker from 46 to \SI{83}{\decibelA} in \dba{3} intervals using a custom automated procedure in a soundproof box \citep{Ooi2021b}, at a distance of \SI{1}{\meter} from a measurement microphone (146AE, G.R.A.S. Sound \& Vibration A/S, Holte, Denmark). The desired output sound level of the masker corresponding to the gain value determined by the AMSS was achieved by energetic interpolation and compensation for distance (inverse square law) and number of speakers (4 speakers).

A total of 481 instances of maskers selected by the AMSS and reproduced over the loudspeakers at the \rtgp\ were logged across 18 of the 20 sessions during the AMSS condition. Maskers \texttt{bird\_00069} (\SI{26}{\percent}) and \texttt{bird\_00075} (\SI{67}{\percent}) were selected the most often, which were sometimes interjected by \texttt{bird\_00071} (\SI{5.8}{\percent}), \texttt{bird\_00025} (\SI{1.0}{\percent}) and \texttt{bird\_00012} (\SI{0.2}{\percent}), as delineated in \Cref{tab:predictions}. The frequency of masker presentation in \Cref{tab:predictions} depicts an average participant's exposure during the \SI{10}{\minute} listening period preceding the evaluation at the \rtgp\ for the \amss\ group. 

\input{tables/predictions}

\subsection{Non-acoustic environmental conditions for in-situ validation study}

As measured by the instruments shown in \Cref{tab:instrument}, the in-situ experimental conditions exhibited notable stability across all parameters for both the \amss\ and \amb\ groups, as presented in \Cref{tab:enviro}. The prevailing temperature and humidity levels align with the characteristic hot and humid tropical climate of \db{Singapore}, complemented by wind speeds indicative of light air. A noteworthy consideration is the absolute luminance levels, which were damped by the tinting on the protective cover over the sensor.

Importantly, the air quality remained within healthy limits throughout the entire study duration. Employing Wilcoxon rank-sum tests at a \SI{5}{\percent} significance level revealed no significant distinctions between the \amss\ and \amb\ groups across key environmental parameters of temperature, relative humidity, luminance, wind speed, 24-hour PSI, and PM2.5 readings. Given the inherent in-situ nature of this study, where environmental parameters are beyond direct experimental control, the discovery of non-significant differences between groups is fortuitous but noteworthy. This outcome allays concerns associated with potential confounding factors stemming from divergent environmental conditions between the \amss\ and \amb\ groups.

\input{tables/envirosummary}

\subsection{Participants}

A cohort of \num{70} participants participated in this study. Recruitment was executed through mobile messaging channels and the distribution of advertisements via grassroots organisations. The study inclusion criteria mandated that participants reside within the designated postal sector of the study site (i.e., postal sector \db{82}) and fall within the age range of 21 to 70 years. Participants received remuneration in the form of supermarket vouchers with a value of \db{\$30 (Singapore dollars)}.

Due to the onset of a thunderstorm midway through one of the study sessions, data from the two participants for that session were deemed unreliable and subsequently excluded from the analysis. The final dataset comprised responses from 68 participants, consisting of \num{40} females (\SI{59}{\percent}) and \num{28} males (\SI{41}{\percent}), with a mean age of 41.75 and standard deviation of age of 12.83, as detailed in \Cref{tab:demographic}. Participants were generally working-class adults, but the employment status varied among individual participants, with a majority being employed (\SI{71}{\percent}), followed by retirees (\SI{8.8}{\percent}), students (\SI{8.8}{\percent}), unemployed individuals (\SI{4.4}{\percent}), and a segment that either did not disclose or fell into the ``other'' category (\SI{7.4}{\percent}).  
The \amss\ group comprised 36 participants, while the \amb\ group consisted of 32 participants. Variations in proportions of gender between the \amss\ and \amb\ groups were determined to be non-significant via four-sample test for equality of proportions without continuity correction ($p=0.09$). On the other hand, the age was similarly distributed between \amss\ and \amb ($p=0.91$). The central tendencies and dispersion of the self-assessed PSS-10, INS, WHO-5, and baseline annoyance across all noise categories were similar across both \amss\ and \amb\ groups, as detailed in \Cref{tab:demographic}.

Tests of distribution equality were performed across age, PSS-10, INS, WHO-5, and all baseline annoyance categories, acknowledging the potential influence of non-acoustical factors on soundscape perception \citep{gao_effects_2023,Tarlao2020nonacoustic}. Analysis using the exact two-sample Kolmogorov-Smirnov test revealed no significant differences between the \amss\ and \amb\ groups, as listed in the $p$-value column in \Cref{tab:demographic}.

\input{tables/demographic}

\subsection{Data analysis}\label{sec:data_analysis}

From the binaural recordings collected in \Cref{sec:experimental_design}, objective acoustic and psychoacoustic indices were computed with a commercial software package (ArtemiS suite, HEAD acoustics GmbH, Herzogenrath, Germany) on the representative channel with the highest value \citep{iso12913-3}. These included both the A- and C-weighted equivalent sound pressure level over each \SI{10}{\minute} listening period ($L_\text{A,eq}$; $L_\text{C,eq}$), and the \SI{95}{\percent} exceedance level of psychoacoustic loudness ($N_\text{95}$) as computed with ISO 532-1 \citep{InternationalOrganizationforStandardization2017a}. Whereas the $L_\text{A,eq}$ and $L_\text{C,eq}$ metrics are commonly used in noise policies, the $N_\text{95}$ was previously found to correlate strongly with the perceived loudness of traffic sounds \citep{Lam2023}.

For consistency and comparability, the scales for all items in the site evaluation questionnaire were normalized such that all values ranged from $-1$ to 1 before further analysis was performed. The PAQ items were also transformed into the normalized quantities ``ISO Pleasantness (\isopl)'' and ``ISO Eventfulness (\isoev)'' based on the definition given in ISO 12913-3. Specifically, we computed \isopl\ and \isoev\ as
\begin{align}
	\textit{ISOPL} &= \frac{2\left(r_\text{pl}-r_\text{an}\right)+\sqrt{2}\left(r_\text{ca}-r_\text{ch}+r_{\text{vi}}-r_{\text{mo}}\right)}{8+8\sqrt{2}} \in [-1,1], \text{\hspace{2mm}and} \label{eq:ISOPl_general} \\
	\textit{ISOEV} &= \frac{2\left(r_\text{ev}-r_\text{un}\right)+\sqrt{2}\left(r_{\text{ch}}-r_{\text{ca}}+r_{\text{vi}}-r_{\text{mo}}\right)}{8+8\sqrt{2}} \in [-1,1], \label{eq:ISOEv_general}
\end{align}
\noindent where $r_{\text{pl}}, r_{\text{ev}}, r_{\text{ch}}, r_{\text{vi}}, r_{\text{un}}, r_{\text{ca}}, r_{\text{an}}, r_{\text{mo}} \in \{1,2,\dots,5\}$ are the extent to which the soundscape was respectively perceived to be \textit{pleasant}, \textit{eventful}, \textit{chaotic}, \textit{vibrant}, \textit{uneventful}, \textit{calm}, \textit{annoying}, and \textit{monotonous}, on a scale of 1 to 5. Separate positive affect (\pa) and negative affect (\na) scores were also computed from the responses to the I-PANAS-SF, as recommended by \cite{Watson1988}.

In the scope of a between-within experimental design, quantitative attributes were assessed using a two-way linear mixed effects with a repeated measures approach. The factor within subjects, termed \textit{site}, featured two levels: \gfp\ and \rtgp. Simultaneously, the between-subject factor, termed \textit{condition}, featured two levels: \amb\ and \amss. 

For the examination of the attributes in \noi, \nat, \hum, \na, \osq, \appr, \pln\}, a non-parametric two-way linear mixed effects repeated measures type III rank-transformed analysis of variance (\rtrmanova) was applied. This method involves replacing the original data with their ranks, a technique well suited for multiple comparisons \citep{Conover1981}. The model included a random intercept to account for potential variability in baseline responses across participants.

We utilized a similar analytical approach to investigate the derived attributes in \{\pa, \isopl, \isoev, \fas, \ba, \com, \ec, \es\}, namely a non-parametric two-way linear mixed effects repeated measures type III analysis of variance (\rmanova). Notably, we refrained from rank transformation in this case, because the residuals exhibited normality as confirmed through Shapiro-Wilk's test ($p>0.05$).

In addition, to assess the potential impact of order effects and group sizes, multiple comparisons were made across all the attributes of soundscape evaluation in \{\noi, \nat, \hum, \na, \pa, \osq, \appr, \pln, \fas, \ba, \com, \ec, \es, \isopl, \isoev\} for each condition, employing the non-parametric two-sample Kolmogorov-Smirnov (KS) test. For the analysis of order effects, the responses were grouped into a sample from all participants who evaluated the \gfp\ first followed by the \rtgp, and another sample from all participants who evaluated the \rtgp\ first followed by the \gfp. For the analysis of group sizes, the responses were grouped into a sample from all participants who evaluated the sites by themselves, and another sample from all participants who evaluated the sites with at least one other participant in the same session. To mitigate false discovery rates due to multiple comparisons, $p$-values were adjusted using the Benjamini-Hochberg (BH) method separately for each condition. 

All data analyses were conducted with the R programming language (Version 4.3.1; \citet{RCoreTeam2023}) on a 64-bit ARM environment. Specifically, the analyses were performed with the following packages: KS test, BH correction, Shapiro-Wilk Normality Test with \texttt{stats} (Version 4.3.1; \citet{RCoreTeam2023}); \rmanova\ and \rtrmanova\ with \texttt{lmerTest} (Version 3.1.3; \citet{Kuznetsova2017}) and \texttt{car} (Version 3.1.2; \citet{Fox2019car}); Omega effect size with \texttt{effectsize} (Version 0.8.3; \citet{Ben-Shachar2020}); and contrast tests with \texttt{emmeans} (Version 1.8.7; \citet{emmeans2023}).

%% file: tables/instruments.tex
\begin{table}[ht]
\centering
\caption{Critical specifications of measurement instruments.}
\label{tab:instrument}

\begin{tabularx}{\linewidth}{%
>{\raggedright\arraybackslash}p{3.5cm}%
>{\raggedright\arraybackslash}p{3cm}%
>{\raggedright\arraybackslash}p{2cm}%
>{\raggedright\arraybackslash}p{3cm}X%
}
\toprule
\textbf{Instrument}
&  \textbf{Metric}
&  \textbf{Range}
&  \textbf{Accuracy}
&  \textbf{Resolution/Sensitivity}\\

\midrule
\multirow{2}{3.5cm}{TYPE 4101-B Binaural Microphone}
& Sound pressure (\si{\pascal})
& \SI{20}{\hertz} -- \SI{5}{\kilo\hertz}

& $\pm \SI{2}{\decibel}$ re \SI{1}{\kilo\hertz}
& 20 mV/Pa $\pm \SI{3}{\decibel}$\\

&
& 5 -- \SI{20}{\kilo\hertz}
& \SI{3}{\decibel} soft boost at \SI{0}{\degree} incidence
& \\

\midrule
\multirow{2}{3.5cm}{BME280 Digital humidity, pressure and temperature sensor}
&  Temperature (\si{\degreeCelsius})
&  0 -- 65
&  $\pm$0.5
& 0.01\\
         
&  Relative humidity (\%RH)
&  0 -- 100 \newline (0 -- 60\si{\degreeCelsius})
&  $\pm$3
& 0.008\\

\midrule
LTR-559ALS-01 Optical Sensor
&  Luminance (\si{\lux})
&  1
&  64000
& 0.977\\
\bottomrule
\end{tabularx}
\end{table}

%% file: tables/predictions.tex
\setlength{\LTpost}{0mm}
\begin{table}[ht]
\centering
\caption{Frequency distribution of the maskers chosen by the AMSS during the 10-\si{\minute} listening period across all ``AMSS'' group participants. Description and availability of the corresponding maskers as detailed by \citet{Ooi2023a} in the ARAUS dataset.}
\label{tab:predictions}

\begin{tabularx}{0.68\linewidth}{XcX}
\toprule

\textbf{Maskers} 
& \textbf{Frequency (\%)}
& \textbf{Description} \\ 

\midrule\addlinespace[2.5pt]

bird\_00012 
& 0.2\% 
& Bahama Mockingbird\textsuperscript{\textit{a}}\\ 

bird\_00025 
& 1.0\% 
& Baltimore Oriole\textsuperscript{\textit{b}}\\ 

bird\_00069 
& 26\% 
& Northern Cardinal\textsuperscript{\textit{c}}\\ 

bird\_00071 
& 5.8\% 
& Veery\textsuperscript{\textit{d}}\\ 

bird\_00075 
& 67\% 
& Common Redshank\textsuperscript{\textit{e}}\\ 

\bottomrule

\end{tabularx}
\begin{minipage}{0.68\linewidth}
\textsuperscript{\textit{a}}Paul Driver, XC140239. Accessible at \url{www.xeno-canto.org/140239}.\\
\textsuperscript{\textit{b}}Eric DeFonso, XC370500. Accessible at \url{www.xeno-canto.org/370500}.\\
\textsuperscript{\textit{c}}Christopher McPherson, XC601752. Accessible at \url{www.xeno-canto.org/601752}.\\
\textsuperscript{\textit{d}}Christopher McPherson, XC602571. Accessible at \url{www.xeno-canto.org/602571}.\\
\textsuperscript{\textit{e}}Joao Tomas, XC604437. Accessible at \url{www.xeno-canto.org/604437}.\\
\end{minipage}
\end{table}

%% file: tables/envirosummary.tex
\setlength{\LTpost}{0mm}
\begin{table}[ht]
\centering
\caption{Summary statistics of environmental parameters captured at \chreplaced[comment=R1.2 R2.4]{\rtgp}{RTGP} during the 10-\si{\minute} listening period across all participants.}
\label{tab:enviro}

\begin{tabularx}{0.7\linewidth}{l%
>{\raggedright\arraybackslash}p{2.5cm}%
>{\raggedright\arraybackslash}p{2.5cm}%
X}
\toprule
\textbf{Environmental Parameter} 
& \textbf{AMSS}\textsuperscript{\textit{1}} 
& \textbf{Ambient}\textsuperscript{\textit{1}} 
& \textbf{p-value}\textsuperscript{\textit{2}} \\

\midrule\addlinespace[2.5pt]
Temperature (\si{\degreeCelsius}) 
& 31.64 (1.37) 
& 33.39 (2.13) 
& 0.083 \\

Relative Humidity (\%RH)
& 59.09 (4.20) 
& 56.02 (6.96) 
& 0.494 \\ 

Luminance (\si{\lux}) 
& 314.65 (132.44) 
& 334.45 (162.26) 
& 0.750 \\  

Wind Speed (\si{\kilo\meter\per\hour}) 
& 3.63 (0.76) 
& 3.17 (1.25) 
& 0.259 \\  

24-\si{\hour} PSI 
& 50.17 (6.08) 
& 45.50 (6.50) 
& 0.203 \\ 

PM2.5 (\si{\micro\gram\per\meter\cubed})
& 16.00 (4.15) 
& 11.92 (4.54) 
& 0.098 \\

\bottomrule
\end{tabularx}
\begin{minipage}{0.7\linewidth}
\textsuperscript{\textit{1}}Reported as ``Mean (Standard deviation)''\\
\textsuperscript{\textit{2}}Wilcoxon rank sum exact test; Wilcoxon rank sum test\\
\end{minipage}
\end{table}

%% file: tables/demographic.tex
\setlength{\LTpost}{0mm}
\begin{table}[ht]

\caption{Summary of participant demographics and non-acoustic factors (PSS-10, WNSS, WHO-5, baseline annoyance) across each condition (\amss\ and \amb).} \label{tab:demographic}

\begin{tabularx}{1\linewidth}{X%
*{3}{>{\raggedleft\arraybackslash}p{8em}}%
>{\raggedleft\arraybackslash}p{5em}
}
\toprule
& \textbf{Overall}, N = 68\textsuperscript{\textit{1}} 
& \textbf{Ambient}, N = 32\textsuperscript{\textit{1}} 
& \textbf{AMSS}, N = 36\textsuperscript{\textit{1}} & \textbf{p-value}\textsuperscript{\textit{2}} \\

\midrule\addlinespace[2.5pt]
\textbf{Gender} &  &  &  & 0.09 \\ 
\hspace{2em}Female & 40 (59\%) & 21 (66\%) & 19 (53\%) &  \\
\hspace{2em}Male & 28 (41\%) & 11 (34\%) & 17 (47\%) &  \\  
\textbf{Age} & 41.75 (12.83) & 42.00 (13.22) & 41.53 (12.65) & 0.91 \\ 
\textbf{Occupation} &  &  &  &\\ 
\hspace{2em}Employed & 48 (71\%) & 26 (72\%) & 22 (69\%) &\\ 
\hspace{2em}Other & 1 (1.5\%) & 1 (2.8\%) & 0 (0\%) &\\ 
\hspace{2em}Rather not say & 4 (5.9\%) & 2 (5.6\%) & 2 (6.3\%) &\\ 
\hspace{2em}Retired & 6 (8.8\%) & 2 (5.6\%) & 4 (13\%) &\\ 
\hspace{2em}Student & 6 (8.8\%) & 3 (8.3\%) & 3 (9.4\%) &\\ 
\hspace{2em}Unemployed & 3 (4.4\%) & 2 (5.6\%) & 1 (3.1\%) &\\ 

\textbf{PSS-10} 
& 0.51 (0.13) & 0.51 (0.13) & 0.51 (0.14) & 0.94 \\  
\textbf{INS}  
& 0.67 (0.06) & 0.67 (0.05) & 0.67 (0.06) & 0.72 \\
\textbf{WHO-5} 
& 0.62 (0.17) & 0.59 (0.17) & 0.65 (0.16) & 0.54 \\
\textbf{BA\textsubscript{aircraft}} & 3.93 (1.39) & 3.88 (1.41) & 3.97 (1.38) & 0.82 \\
\textbf{BA\textsubscript{mrt}} & 2.35 (1.22) & 2.59 (1.29) & 2.14 (1.13) & 0.46 \\ 
\textbf{BA\textsubscript{consite}} & 3.53 (1.30) & 3.59 (1.29) & 3.47 (1.32) & 0.80 \\ 
\textbf{BA\textsubscript{reno}} & 3.46 (1.34) & 3.59 (1.39) & 3.33 (1.31) & 0.59 \\ 
\textbf{BA\textsubscript{traffic}} & 3.46 (1.20) & 3.53 (1.14) & 3.39 (1.27) & 0.90 \\ 
\textbf{BA\textsubscript{animals}} & 2.12 (1.10) & 1.94 (1.05) & 2.28 (1.14) & 0.28 \\ 
\textbf{BA\textsubscript{children}} & 2.51 (1.17) & 2.66 (1.21) & 2.39 (1.13) & 0.51 \\
\textbf{BA\textsubscript{people}} & 2.34 (1.02) & 2.47 (1.05) & 2.22 (0.99) & 0.28 \\
\textbf{BA\textsubscript{others}} & 2.35 (1.18) & 2.38 (1.10) & 2.33 (1.26) & 0.83 \\
\bottomrule

\end{tabularx}
\begin{minipage}{1\linewidth}
\textsuperscript{\textit{1}}Gender and occupation reported as ``Count (\%)''; all others reported as ``Mean (Standard deviation)'';
\textsuperscript{\textit{2}}Four-sample test for equality of proportions without continuity correction for gender, and Exact two-sample Kolmogorov-Smirnov test otherwise\\
\end{minipage}

\end{table}

%% file: body/03results.tex


\section{Results: Site evaluation questionnaire}\label{sec:SEQ}

\begin{figure}
    \centering
    \includegraphics[width=0.9\textwidth]{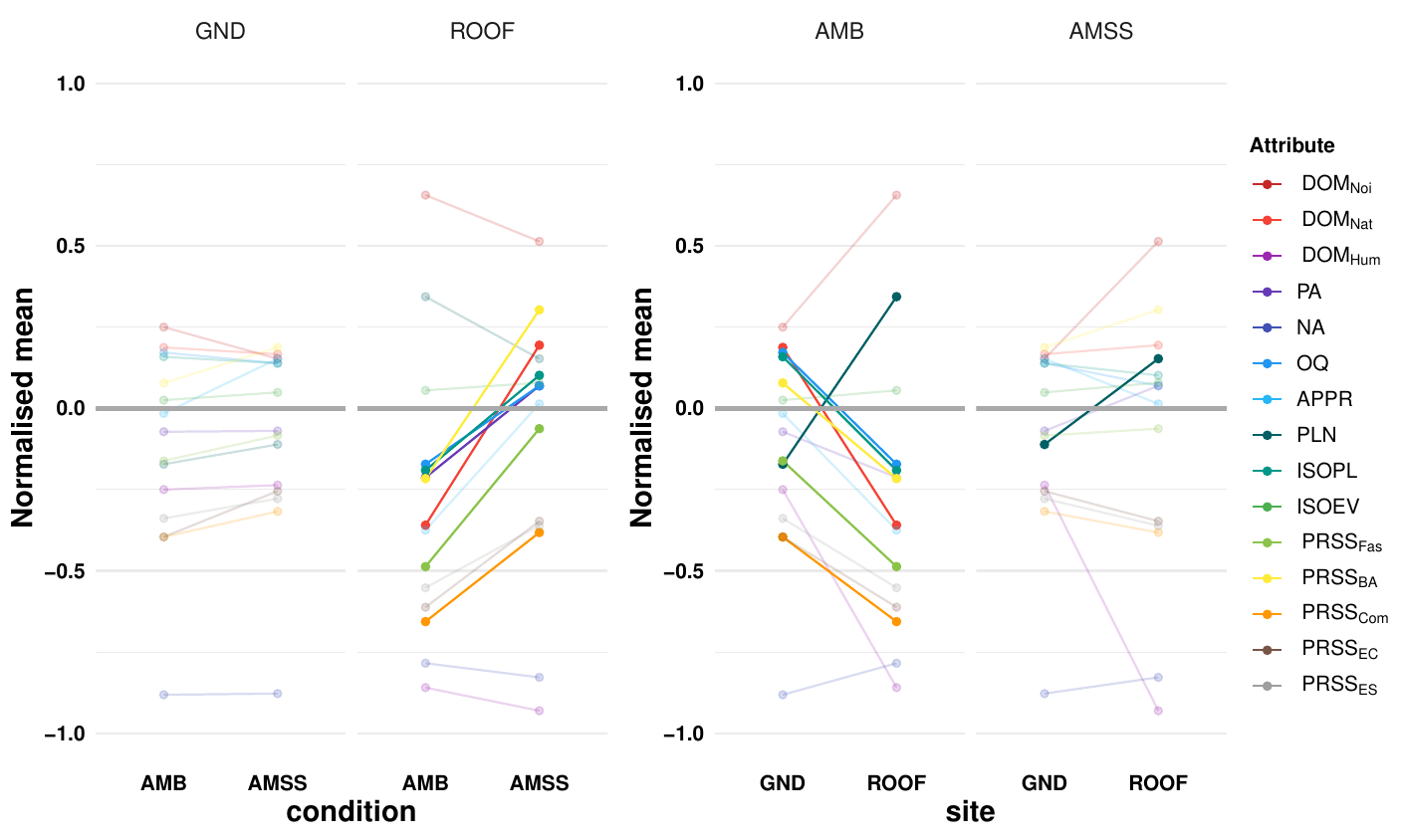}
    \caption{Simple contrast of means across all perceptual attributes organized by \textit{condition} and \textit{site}. Contrasts by \textit{condition} are between group at each \textit{site}, whereas contrasts by \textit{site} are within group for each \textit{condition}. The scales for all attributes are normalised to the range [$-$1,1]. Significant differences as determined by posthoc contrast tests are accentuated}
    \label{fig:contrast}
\end{figure}

A summary of the mean $\mu$ and standard deviation $\sigma$ of quantities derived from the site evaluation questionnaire is shown in \Cref{tab:meanpaq}. As mentioned in \Cref{sec:data_analysis}, all quantities are normalized to the same range $[-1,1]$ for the presentation of results in this section. Furthermore, only significant results of the \rtrmanova\ and \rmanova\ are presented here, but full details of the tests are given in \ref{sec:statresults}, \Cref{tab:statresults}. For clarity, the scale changes are illustrated in \Cref{fig:contrast}, similarly organised by \textit{site} and \textit{condition}, where the significant posthoc contrast pairs are accentuated. 

\input{tables/attributes_mean_sd}

\subsection{Contrast by \textit{condition} between groups at each \textit{site}} \label{sec:bycond}

At \gfp, no significant interaction effects were noted across the perceptual metrics between \amss\ and \amb\ groups (Figure~\ref{fig:contrast}, leftmost), aligning with expectations given the absence of AMSS at \gfp. Consistency in perception among \amss\ and \amb\ groups suggests stability in the \gfp\ soundscape and uniform participant perceptions, facilitating comparison at \rtgp.

At \rtgp, the AMSS induced significant improvements in \chadded[comment=R2.1]{Pleasantness (\isopl), dominance of natural sounds (\nat), overall soundscape quality (\osq), positive affect (\pa), \textit{Fascination} dimension of PRSS (\fas), \textit{Being-Away} dimension of PRSS (\ba), and \textit{Compatibility} dimension of PRSS (\com)} of the traffic-exposed soundscape (Figure~\ref{fig:contrast}, second from left). Notably, a \SI{14.62}{\percent} increase in \isopl\ marks a key ``positive transition'' from a ``bad'' ($\mu=-0.19$) to a ``good'' ($\mu=0.10$) soundscape, validating the efficacy of AMSS in improving PAQ. The \isoev, a PAQ measure of a soundscape's ``Eventfulness''  \citep{iso12913-3}, was unaffected by the AMSS intervention, as desired.

Though traffic noise dominance (\noi) decreased insignificantly by \SI{7.12}{\percent}, a more-than-proportionate \SI{27.69}{\percent} positive transition in natural sound dominance (\nat) from \amb\ ($\mu=-0.36$) to \amss\ ($\mu=0.19$) was observed at the \rtgp. Significant \SI{12.07}{\percent} positive transition from \amb\ ($\mu=-0.17$) to \amss\ ($\mu=0.07$) was also observed in \osq, but increased appropriateness of the soundscape (\appr; \SI{19.44}{\percent}) and decreased perceived loudness (\pln; \SI{-9.55}{\percent}) were not significant with the AMSS intervention at \rtgp. AMSS significantly increased positive affect (\pa; \SI{14.10}{\percent}) \chadded[comment=R2.5]{, suggesting an increase in positive emotions such as \textit{attentive}}, as measured by the I-PANAS-SF. but did not significantly decrease negative affect (\na; \SI{-2.17}{\percent}), \chadded[]{which refer to negative emotions like \textit{nervous}}.

The restorative potential of the AMSS was evidenced by significant improvements with AMSS at the \rtgp\ in PRSS dimensions of \textit{Fascination} (\SI{21.22}{\percent}), \textit{Being-Away} (\SI{25.97}{\percent}) and \textit{Compatibility} (\SI{13.72}{\percent}). Particularly, a positive transition was observed in \ba\ from \amb\ ($\mu=-0.22$) to \amss\ ($\mu=0.30$), which is an indicator of respite provided by the soundscape from daily stressors \citep{Payne2013,Payne2018}. However, improvements in \textit{Extent} sub-dimensions of \ec\ (\SI{13.24}{\percent}) and \es\ (\SI{9.55}{\percent}) were not significant.

\subsection{Contrast by \textit{sites} within group under each \textit{condition}} \label{sec:bysite}

Under the \amb\ \textit{condition}, which is indicative of the difference between the \textit{sites} before intervention, significant changes were noted in \isopl, \nat, \osq, \pln, \fas, \ba, and \com\ (Figure~\ref{fig:contrast}, second from right). The PAQ in terms of \isopl\ was rated a significant \SI{17.47}{\percent} lower at the \rtgp\ ($\mu=-0.19$) than at the \gfp\ ($\mu=0.16$), whereas \isoev\ was equally neutral between the \gfp\ ($\mu=0.03$) and pre-intervention \rtgp\ ($\mu=0.06$) \textit{sites}. 

As expected but not significant, traffic noise dominance was \SI{20.31}{\percent} higher at the traffic-exposed \rtgp\ ($\mu=0.66$) than at the \gfp\ ($\mu=0.25$). Additionally, human sounds were \SI{30.47}{\percent} more dominant at \gfp\ ($\mu=-0.25$) than at the almost non-existent levels at the \rtgp\ ($\mu=-0.86$). On the other hand, natural sounds were scarce ($\mu=-0.36$) and a significant \SI{27.34}{\percent} lower at \rtgp\ than \gfp\ ($\mu=-0.19$).

Before intervention, the \osq\ was poor at the \rtgp\ ($\mu=-0.17$) and a significant \SI{17.19}{\percent} lower than the \osq\ of the \gfp\ ($\mu=0.17$). Similarly but not significantly, \rtgp\ was rated \SI{17.97}{\percent} less appropriate than the \gfp. Interestingly, no significant changes in positive (\pa) or negative (\na) affect, were observed between \gfp\ and \rtgp\ without intervention. 

Regarding restorative indicators, significant differences were noted in dimensions such as \fas, \ba, and \com, indicating poorer restorativeness at the \rtgp\ compared to \gfp. Notably, the restorativeness of \gfp\ was only slightly conducive in terms of \ba\ ($\mu=0.08$).

Under the \amss\ \textit{condition}, no significant changes were found between \gfp\ and \rtgp\ sites, except for \pln\ (Figure~\ref{fig:contrast}, rightmost). This suggests that the AMSS effectively improved the \isopl, \nat, \osq, \fas, \ba, and \com\ scores of the \rtgp\ similar to those at the traffic-shielded \gfp. Although perceived loudness increased (\SI{13.19}{\percent}), it was to a lesser extent than without AMSS intervention (\SI{25.78}{\percent}).

\subsection{Correlation between subjective metrics}

Based on the Holm-adjusted Kendall correlation, listed in \Cref{tab:kendall_corrmat}, \isopl\ was found to be significantly positively correlated with \nat\ ($\tau=0.2887$), \osq\ ($\tau=0.6204$), \appr\ ($\tau=0.5048$), and with the restorative metrics of \fas\ ($\tau=0.3389$), \ba\ ($\tau=0.4894$), \com\ ($\tau=0.5367$), and \ec\ ($\tau=0.5215$). In contrast, \isopl\ was negatively correlated with \noi\ ($\tau=-0.3028$), \na\ ($\tau=-0.2885$), and \pln\ ($\tau=-0.3901$), but was not significantly associated with \hum, \pa, \isoev, and \es.

\input{tables/corrmatrix}

\subsection{Effect of order, group size and initial conditions}

The KS tests with BH adjustments across each condition (\amss\ and \amb) demonstrated that none of the attributes from the site evaluation questionnaire were influenced by the order in which the participants assessed the sites (\gfp$\rightarrow$\rtgp\ or \rtgp$\rightarrow$\gfp), as well as the number of participants in each session ($1$ or $>1$). In other words, the results of this study were not subject to potentially confounding order effects and the possibility of participants affecting each others' responses to the soundscapes experienced. Full details of the results can be found in \ref{sec:statresults}, \Cref{tab:kstest}.

Posthoc contrast tests on \pa\ between \amss\ and \amb\ groups at the meeting point and the absence of interaction effects on \na\ revealed no significant differences between the \amss\ and \amb\ groups in terms of positive and negative affect states before the commencement of the experiment.

\section{Results: Objective binaural measurements}\label{sec:obj}

\begin{figure}
    \centering
    \includegraphics[width=0.95\linewidth]{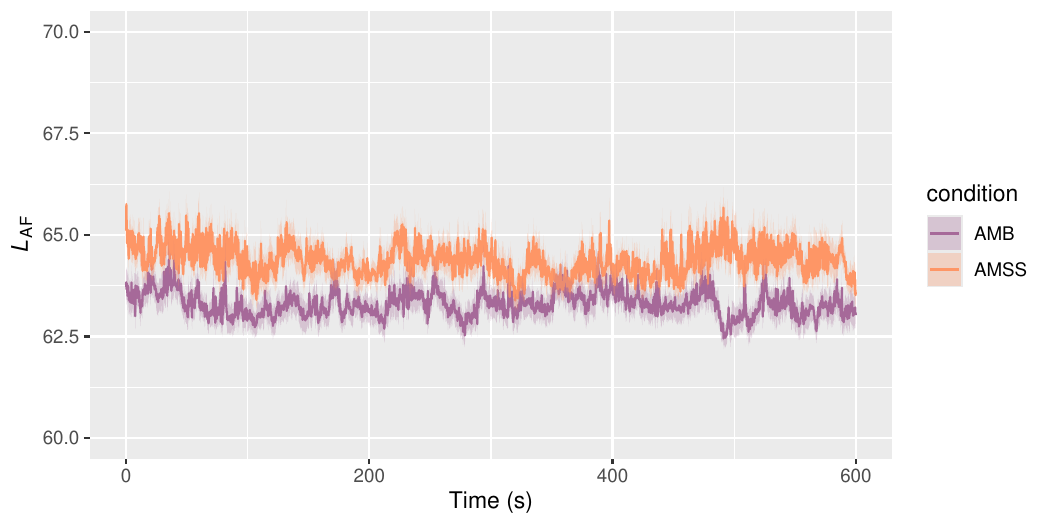}
    \caption{Energetic mean A-weighted, fast time-weighted sound pressure level, $L_{AF}$, of the loudest binarual channel across all sessions in the AMSS and AMB groups at \rtgp\ without aircraft flyby. The shaded error envelope represents the standard error of the mean.}
    \label{fig:spltime}
\end{figure}

At the \rtgp, the mean 10-\si{\minute} \laeq\ was $64.97 \pm 3.38$ \dba{}\ for the \amss\ group and $63.96 \pm 2.95$ \dba{}\ for the \amb\ group, as shown in \Cref{tab:isopl_obj}. Since AMSS was active for the \amss\ group, it caused a slight but imperceptible increase (about \dba{1}) in mean SPL over the study duration at the \rtgp. This suggests that on average, AMSS selected masker gains that were well below the ambient SPLs. For instance, if AMSS reproduced maskers at the same SPL as the ambient acoustics, it would result in a \dba{3} increase. This difference is further reduced to less than \dba{1}\ when one of the \amss\ sessions affected by aircraft noise was omitted from the computed mean. \chadded[comment=R2.2]{With the removal of sessions affected by aircraft flybys, the small standard deviation of \dba{1.07} further indicates that the SPL remained relatively consistent across the \SI{10}{\minute} listening period in all sessions, as shown in \Cref{fig:spltime}. The slight elevation in SPL due to the AMSS appeared evenly distributed throughout the entire listening period across all session.}

At the \gfp, in contrast, the mean 10-\si{\minute} \laeq\ was $63.78 \pm 7.17$ \dba{}\ for the \amss\ group and $57.91 \pm 1.46$ \dba{}\ for the \amb\ group. The relatively higher mean SPL and standard deviation of SPL at the \chadded[comment=R1.2]{\gfp\ }for the \amss\ group was due to aircraft flybys occurring in three of sessions at the \gfp\ and one at \rtgp, which when omitted from the computation of the mean, would have given an \laeq\ of $58.26 \pm 1.77$ \dba{}\ at the \gfp\ and $64.25 \pm 1.07$ \dba{}\ at the \rtgp\ instead. Hence, the difference in \laeq\ between the \gfp\ and \rtgp\ was about \dba{6}\ in both \amb\ and \amss\ groups. A similar trend was observed in the C-weighted equivalent sound pressure level, \lceq\ and in \nnf, where the differences between the \textit{sites} were about 3 to \dbc{5}\ and about 5 to 6 soneGF, respectively.  

To examine the relationship between objective (psycho)acoustic parameters, and soundscape and restorative indicators, a correlation and distribution analysis was conducted between objective parameters (\laeq, \lceq, \nnf), and soundscape (\isopl, \osq) and restorative (\fas, \ba, \com) indices that show statistical difference between \amb--\rtgp\ and \amb--\gfp\ in \Cref{sec:SEQ}. The Holm-adjusted Kendall correlation revealed no significant relationships between the (psycho)acoustic parameters and all the soundscape and restorative indices (Table~\ref{tab:kendall_corrmatobj}).

\begin{figure}
    \centering
    \includegraphics[width=1\textwidth]{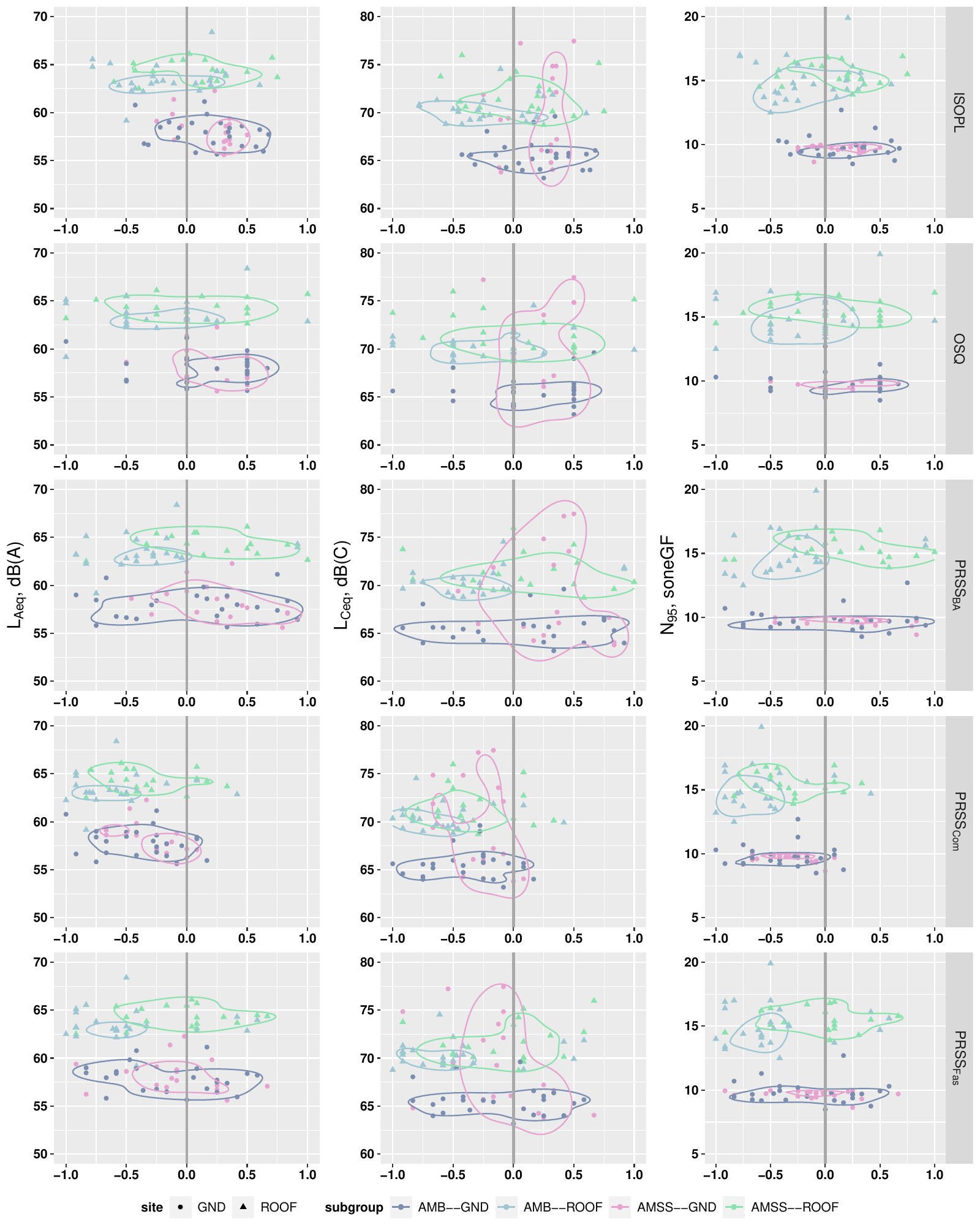}
    \caption{Mean perceptual \isopl, \osq, \fas, \ba, and \com\ scores across all participants per session (y-axis) as a function of normalized objective \laeq, \lceq, and \nnf\  scores of each session (x-axis). Fifty percent of the sessions lie within the median contours computed for \amb--\gfp, \amb--\rtgp, \amss--\gfp, \amss--\rtgp\ contrast subgroups. The left to right columns represent \laeq, \lceq, and \nnf, and each row represent each of the perceptual metrics, respectively.}
    \label{fig:subj_vs_obj}
\end{figure}

The disassociation between objective and perceptual indicators is further illustrated in the median contour plots of the mean perceptual score for each session as a function of each (psycho)acoustic parameter, organised into the \textit{condition}--\textit{site} pairs in \Cref{fig:subj_vs_obj}. Notably, distinct positive shift in median contours across all perceptual indicators was achieved with the introduction of \amss\ at the \rtgp\ despite a similar levels of \laeq, \lceq, or \nnf\ in the \amb--\rtgp\ subgroup. Moreover, \amss--\rtgp\ exhibited similar \isopl, \fas, \ba, and \com\ distributions as both \gfp\ subgroups. Although the \osq\ contours in \amss--\rtgp\ largely overlapped with \amb--\rtgp\ across all objective indices, there is a notable positive shift in the population distribution, as shown in \Cref{sec:bycond}. It is worth noting that the \lceq\ distribution was greatly skewed by the dominant low-frequency content of aircraft flyby sounds in the \amss--\gfp\ sessions, which was not reflected in the \laeq\ and \nnf.

\noindent \input{tables/isopl_obj}

%% file: tables/attributes_mean_sd.tex
\setlength{\LTpost}{0mm}
\setlength{\tabcolsep}{0.16em}
\begin{table}[ht]
\centering
\caption{Mean responses $\mu$ (standard deviation $\sigma$) of perceptual attributes in the site evaluation questionnaire investigated for the validation study, organized by \textit{site} and \textit{condition}. The scales for all attributes are normalised to the range [$-$1,1]. Percentage changes are computed between the \amb\ and \amss\ for \textit{site}, and between \rtgp\ and \gfp\ for \textit{condition} as scale changes on the [$-$1,1] range with respect to the former. For instance, a change from \num{-0.25} in the \amb\ condition to 0.75 in the \amss\ condition would be reported as a \SI{50}{\percent} change. Significant changes as determined by posthoc tests are indicated in bold.}
\label{tab:meanpaq}
\def\arraystretch{1.15} 
\scriptsize

\begin{tabularx}{1\linewidth}{
>{\raggedright\arraybackslash}X%
*{2}{>{\raggedleft\arraybackslash}p{0.08\linewidth}}%
>{\raggedleft\arraybackslash}p{0.05\linewidth}%
*{2}{>{\raggedleft\arraybackslash}p{0.08\linewidth}}%
>{\raggedleft\arraybackslash}p{0.05\linewidth}%
*{2}{>{\raggedleft\arraybackslash}p{0.08\linewidth}}%
>{\raggedleft\arraybackslash}p{0.05\linewidth}%
*{2}{>{\raggedleft\arraybackslash}p{0.08\linewidth}}%
>{\raggedleft\arraybackslash}p{0.05\linewidth}}
\toprule
& \multicolumn{6}{c}{\textit{site}} 
& \multicolumn{6}{c}{\textit{condition}}\\
\cmidrule(l{0.2em}r{0.2em}){2-7} \cmidrule(l{0.2em}r{0.2em}){8-13} 

& \multicolumn{3}{c}{\gfp} 
& \multicolumn{3}{c}{\rtgp} 
& \multicolumn{3}{c}{\amb} 
& \multicolumn{3}{c}{\amss} \\

\cmidrule(l{0.2em}r{0.2em}){2-4} \cmidrule(l{0.2em}r{0.2em}){5-7}
\cmidrule(l{0.2em}r{0.2em}){8-10} \cmidrule(l{0.2em}r{0.2em}){11-13}

& \amb\ 
& \amss\
& $\Delta$(\%) 
& \amb\
& \amss\ 
& $\Delta$(\%) 
& \gfp\  
& \rtgp\
& $\Delta$(\%) 
& \gfp\  
& \rtgp\ 
& $\Delta$(\%) \\

\midrule\addlinespace[2.5pt]

\noi 
& 0.25 (0.44) & 0.15 (0.50) & -4.86 
& 0.66 (0.39) & 0.51 (0.42) & -7.12
& 0.25 (0.44) & 0.66 (0.39) & 20.31 
& 0.15 (0.50) & 0.51 (0.42) & 18.06 \\

\nat
& 0.19 (0.40) & 0.17 (0.49) & -1.04 
& -0.36 (0.50) & 0.19 (0.44) & \textbf{27.69} 
& 0.19 (0.40) & -0.36 (0.50) & \textbf{-27.34}
& 0.17 (0.49) & 0.19 (0.44) & 1.39 \\

\hum
& -0.25 (0.38) & -0.24 (0.60) & 0.69 
& -0.86 (0.34) & -0.93 (0.34) & -3.56 
& -0.25 (0.38) & -0.86 (0.34) & -30.47 
& -0.24 (0.60) & -0.93 (0.34) & -34.72 \\

\pa
& -0.07 (0.43) & -0.07 (0.50) & 0.12 
& -0.21 (0.38) & 0.07 (0.59) & \textbf{14.10} 
& -0.07 (0.43) & -0.21 (0.38) & -7.03 
& -0.07 (0.50) & 0.07 (0.59) & 6.94\\ 

\na
& -0.88 (0.18) & -0.88 (0.23) & 0.17 
& -0.78 (0.30) & -0.83 (0.43) & -2.17 
& -0.88 (0.18) & -0.78 (0.30) & 4.84 
& -0.88 (0.23) & -0.83 (0.43) & 2.50\\

\osq
& 0.17 (0.47) & 0.14 (0.39) & -1.65 
& -0.17 (0.50) & 0.07 (0.55) & \textbf{12.07}
& 0.17 (0.47) & -0.17 (0.50) & -17.19 
& 0.14 (0.39) & 0.07 (0.55) & -3.47 \\

\appr
& -0.02 (0.39) & 0.15 (0.44) & 8.42 
& -0.38 (0.49) & 0.01 (0.57) & 19.44 
& -0.02 (0.39) & -0.38 (0.49) & -17.97 
& 0.15 (0.44) & 0.01 (0.57) & -6.94\\

\pln
& -0.17 (0.35) & -0.11 (0.49) & 3.04 
& 0.34 (0.43) & 0.15 (0.55) & -9.55
& -0.17 (0.35) & 0.34 (0.43) & \textbf{25.78} 
& -0.11 (0.49) & 0.15 (0.55) & \textbf{13.19} \\

\isopl
& 0.16 (0.32) & 0.14 (0.30) & -1.00 
& -0.19 (0.38) & 0.10 (0.45) & \textbf{14.62} 
& 0.16 (0.32) & -0.19 (0.38) & \textbf{-17.47}
& 0.14 (0.30) & 0.10 (0.45) & -1.86\\

\isoev
& 0.03 (0.23) & 0.05 (0.23) & 1.19 
& 0.06 (0.24) & 0.08 (0.26) & 1.18
& 0.03 (0.23) & 0.06 (0.24) & 1.50 
& 0.05 (0.23) & 0.08 (0.26) & 1.49 \\ 


\fas 
& -0.16 (0.44) & -0.08 (0.39) & 3.91
& -0.49 (0.43) & -0.06 (0.50) & \textbf{21.22}
& -0.16 (0.44) & -0.49 (0.43) & \textbf{-16.28 }
& -0.08 (0.39) & -0.06 (0.50) & 1.04 \\ 

\ba
& 0.08 (0.59) & 0.19 (0.48) & 5.47 
& -0.22 (0.50) & 0.30 (0.68) & \textbf{25.97} 
& 0.08 (0.59) & -0.22 (0.50) & \textbf{-14.71} 
& 0.19 (0.48) & 0.30 (0.68) & 5.79\\

\com
& -0.40 (0.35) & -0.32 (0.30) & 3.94 
& -0.66 (0.35) & -0.38 (0.41) & \textbf{13.72} 
& -0.40 (0.35) & -0.66 (0.35) & \textbf{-13.02} 
& -0.32 (0.30) & -0.38 (0.41) & -3.24\\

\ec
& -0.40 (0.33) & -0.25 (0.29) & 7.06 
& -0.61 (0.34) & -0.35 (0.39) & 13.24
& -0.40 (0.33) & -0.61 (0.34) & -10.81 
& -0.25 (0.29) & -0.35 (0.39) & -4.63 \\

\es 
& -0.34 (0.33) & -0.28 (0.32) & 3.04 
& -0.55 (0.29) & -0.36 (0.31) & 9.55
& -0.34 (0.33) & -0.55 (0.29) & -10.68 
& -0.28 (0.32) & -0.36 (0.31) & -4.17\\

\bottomrule
\end{tabularx}
\end{table}

%% file: tables/corrmatrix.tex
\begin{table}[ht]
\scriptsize
\centering
\caption{Kendall correlation matrix between all attributes in the site evaluation questionnaire where the significance of each entry in the upper triangle is denoted with a Holm-adjusted $p$-value and each entry in the lower triangle is denoted with an unadjusted $p$-value. Asterisks indicate $\text{*}p<0.05$; $\text{**}p<0.01$; $\text{***}p<0.001$; $\text{****}p<0.0001$. The unit diagonal has been removed for clarity.}
\label{tab:kendall_corrmat}

\begin{tabularx}{1\linewidth}{
>{\raggedright\arraybackslash}p{4em}%
*{16}{>{\raggedleft\arraybackslash}X}}

\toprule
\multicolumn{1}{l}{} 
& \noi & \hum & \nat & \pa 
& \na  & \osq & \appr & \pln 
& \isopl & \isoev & \fas & \ba
& \com & \ec & \es\\ 

\midrule\addlinespace[2.5pt]
\noi &  &  -.22 &  -.03 &  -.06 &  .06 & **-.35 & *-.30 & ***.44 & *-.30 &  .08 &  -.12 &  -.23 &  -.24 &  -.22 &  -.07 \\ 
\hum & *-.22 &  &  .27 &  .04 &  -.03 &  .08 &  .12 &  -.14 &  .09 &  .01 &  .12 &  .07 &  .07 &  .10 &  .17 \\ 
\nat &  -.03 & **.27 &  &  .18 &  -.11 & *.29 &  .23 &  -.13 & *.29 &  .02 & ..28 &  .24 &  .25 & *.30 & *.29 \\ 
\pa &  -.06 &  .04 & *.18 &  &  -.11 & *.29 &  .23 &  -.02 &  .23 &  .01 & **.35 & ***.39 & **.34 & ***.37 & **.35 \\ 
\na &  .06 &  -.03 &  -.11 &  -.11 &  &  -.22 &  -.20 &  .16 & *-.29 &  .04 &  -.07 &  -.10 &  -.21 &  -.14 &  -.00 \\ 
\osq & ***-.35 &  .08 & ***.29 & ***.29 & **-.22 &  & ***.56 & ***-.47 & ***.62 &  -.13 & **.34 & ***.49 & ***.54 & ***.52 & ..27 \\ 
\appr & ***-.30 &  .12 & **.23 & **.23 & *-.20 & ***.56 &  & ***-.41 & ***.50 &  -.03 & **.35 & ***.43 & ***.48 & ***.47 & ..28 \\ 
\pln & ***.44 & .-.14 &  -.13 &  -.02 & ..16 & ***-.47 & ***-.41 &  & ***-.39 &  .10 &  -.16 &  -.25 & **-.35 & *-.31 &  -.17 \\ 
\isopl & ***-.30 &  .09 & ***.29 & **.23 & ***-.29 & ***.62 & ***.50 & ***-.39 &  &  -.03 & **.34 & ***.50 & ***.52 & ***.46 &  .25 \\ 
\isoev &  .08 &  .01 &  .02 &  .01 &  .04 &  -.13 &  -.03 &  .10 &  -.03 &  &  -.02 &  -.08 &  -.08 &  -.06 &  .03 \\ 
\fas &  -.12 &  .12 & ***.28 & ***.35 &  -.07 & ***.34 & ***.35 & .-.16 & ***.34 &  -.02 &  & ***.61 & ***.57 & ***.55 & ***.65 \\ 
\ba & **-.23 &  .07 & **.24 & ***.39 &  -.10 & ***.49 & ***.43 & **-.25 & ***.50 &  -.08 & ***.61 &  & ***.69 & ***.64 & ***.51 \\ 
\com & **-.24 &  .07 & **.25 & ***.34 & *-.21 & ***.54 & ***.48 & ***-.35 & ***.52 &  -.08 & ***.57 & ***.69 &  & ***.65 & ***.50 \\ 
\ec & **-.22 &  .10 & ***.30 & ***.37 &  -.14 & ***.52 & ***.47 & ***-.31 & ***.46 &  -.06 & ***.55 & ***.64 & ***.65 &  & ***.52 \\ 
\es &  -.07 & ..17 & ***.29 & ***.35 &  -.00 & **.27 & **.28 & *-.17 & **.25 &  .03 & ***.65 & ***.51 & ***.50 & ***.52 &  \\ 

\bottomrule
\end{tabularx}
\end{table}

%% file: tables/isopl_obj.tex
\setlength{\LTpost}{0mm}

\begin{table}[ht]
\centering

\caption{Summary of mean \laeq, \lceq, \nnf, \isopl, \osq, \fas, \ba, and \com values across 20 \amss\ and 24 \amb\ sessions in each of the \gfp\ and \rtgp\ sites. Supplemented mean values for the \amss\ sessions excluding aircraft flyby (3 in \gfp; 1 in \rtgp) are included.}
\label{tab:isopl_obj}

\small
\begin{tabularx}{1\linewidth}{%
>{\raggedright\arraybackslash}p{2.5em}%
*{6}{>{\raggedleft\arraybackslash}X}}
\toprule

& \multicolumn{2}{c}{\textbf{Ambient}}
& \multicolumn{2}{c}{\textbf{AMSS}} 
& \multicolumn{2}{c}{\textbf{AMSS (without aircraft flyby)}} \\
 
\cmidrule(l{0.2em}r{0.2em}){2-3} 
\cmidrule(l{0.2em}r{0.2em}){4-5} 
\cmidrule(l{0.2em}r{0.2em}){6-7} 

& \textbf{GND}, \textit{N} = 24 
& \textbf{ROOF}, \textit{N} = 24 
& \textbf{GND}, \textit{N} = 20 
& \textbf{ROOF}, \textit{N} = 20 
& \textbf{GND}, \textit{N} = 17
& \textbf{ROOF}, \textit{N} = 19 \\

\midrule\addlinespace[2.5pt]
\laeq
& 57.91 (1.46) 
& 63.96 (2.95)
& 61.04 (7.17) 
& 64.97 (3.38) 
& 58.26 (1.77) 
& 64.25 (1.07) \\

\lceq
& 65.60 (1.55) 
& 70.81 (2.54)
& 70.89 (6.42) 
& 72.30 (3.27)
& 68.93 (4.39) 
& 71.71 (2.01) \\

\nnf 
& 9.80 (0.87) 
& 15.03 (1.64)
& 9.67 (0.31) 
& 15.44 (0.87)  
& 9.66 (0.34)
& 15.47 (0.88) \\

\isopl 
& 0.17 (0.32) 
& \num{-0.20} (0.37) 
& 0.17 (0.23) 
& 0.09 (0.38) 
& 0.20 (0.23) 
& 0.08 (0.38) \\

\osq 
& 0.17 (0.29) 
& 0.03 (0.52) 
& 0.14 (0.43) 
& \num{-0.22} (0.47) 
& 0.21 (0.28) 
& 0.00 (0.52) \\

\fas 
& \num{-0.11} (0.38) 
& \num{-0.09} (0.44) 
& \num{-0.13} (0.44) 
& \num{-0.47} (0.44) 
& \num{-0.10} (0.40) 
& \num{-0.10} (0.45) \\

\ba
& 0.21 (0.38) 
& 0.32 (0.63) 
& 0.10 (0.60) 
& \num{-0.21} (0.52) 
& 0.25 (0.37) 
& 0.30 (0.64) \\

\com
& \num{-0.30} (0.24) 
& \num{-0.36} (0.36) 
& \num{-0.38} (0.32) 
& \num{-0.65} (0.37) 
& \num{-0.28} (0.24) 
& \num{-0.37} (0.37) \\

\bottomrule
\end{tabularx}
\begin{minipage}{\linewidth}

\end{minipage}
\end{table}

%% file: body/04discussion.tex
\section{Discussion}
\label{sec:discussion}

For clarity, the research questions put forward in \Cref{sec:research_questions} are discussed sequentially in \Cref{sec:disc_amss_rtgpvsgfp}, \Cref{sec:disc_amss_perceptual}, and \Cref{sec:disc_amss_objective}, respectively. The discussion culminates with the limitations of this study and suggestions for future research in \Cref{sec:limitations}.


\subsection{Assessing perceptual changes brought about by AMSS at the traffic-expose site} \label{sec:disc_amss_rtgpvsgfp}

The lack of studies focusing on \isopl\ as a design goal, especially in the context of augmenting soundscape affected by traffic noise, highlights the novelty of our investigation. However, the findings could be placed in the context of a previous virtual reality-based lab study set in a comparable scenario -- an outdoor recreational space subjected to traffic noise without direct visibility of the traffic source (i.e., location P\textsubscript{2} in \cite{Hong2020b}). In that study, the scale increase in raw \textit{pleasantness} at P\textsubscript{2} in \cite{Hong2020b} ranged from \num{5.00} to \SI{18.33}{\percent} across four types of bird sounds, and from \num{-8.33} to \SI{16.67}{\percent} across four types of water sounds, with each masker augmented \SI{3}{\decibelA} lower than the ambient traffic noise levels at \SI{65.2}{\decibelA}. With a higher increase in raw \textit{pleasantness} of \SI{23.35}{\percent} observed in this in-situ study (\amss--\rtgp: $r_{pl}=0.1389$; \amb--\rtgp: $r_{pl}=-0.3281$), it is reasonable to conclude that the maskers selected by the AMSS indeed prioritize maximizing \isopl, where \textit{pleasantness} is a significant component.

\subsection{Perceptual implications of \isopl\ as a soundscape intervention design goal} \label{sec:disc_amss_perceptual}

While the primary focus of AMSS optimization was \isopl\ enhancement, significant improvements were evident across various soundscape quality and restorative indicators. Notably, the consistent use of birdsongs as maskers led to a significant increase in natural sound dominance (see \Cref{tab:predictions}), correlating with a reduction by \SI{7.1}{\percent} in \hum\ and \SI{3.6}{\percent} in \noi, as explained by the informational masking theory \citep{Kidd2017,Chau2023}.

With the modification of dominant sound source types, AMSS effectively enhanced the overall soundscape quality (\osq) at the traffic-exposed \rtgp, surpassing the \SI{9}{\percent} mean scale increase reported by \cite{VanRenterghem2020} for their manual augmentation approach. While caution is warranted in directly comparing methodologies due to differing environments, the AMSS's autonomous operation suggests a possible advantage over participant-led methods. Notably, the \osq\ contrast in the \amb\ \textit{condition} between \gfp\ and \rtgp, as described in \Cref{sec:bysite}, highlights the substantial impact of traffic noise at the traffic-exposed \rtgp. Additionally, the absence of significant differences between \gfp\ and \rtgp\ within the \amss\ group suggests that AMSS could align the perception of \osq\ at a traffic-exposed \textit{site} with that of the traffic-shielded environment.

The significant positive transition in positive affect (\pa) induced by the AMSS suggests the potential for harnessing the health benefits of natural sounds \citep{Buxton2021}, which is made more accessible through its perception-driven autonomous operation. \chadded[comment=R2.5]{The I-PANAS-SF used in this study is well suited to capture transient emotional states during the soundscape interventions. PANAS, which treats \pa\ and \na\ as independent constructs, provides insight into emotional changes without assuming them as opposites. The observed increase in \pa\ reflects engagement or positive ``activation'' due to AMSS, while the lack of a significant decrease in \na\ does not diminish the positive effects on \pa.} It is also important to note that non-optimized augmentation of natural sounds in urban environments could lead to undesirable effects on mood and affect \citep{Jiang2021,Benfield2010}.

On the contrary, the lack of significant changes in \isoev\ suggests that AMSS did not alter the perceived ``Eventfulness'' of the soundscape. This was likely due to the AMSS's design goal, which focused solely on maximizing \isopl\ without affecting \isoev. Additionally, according to the circumplex model of soundscape perception in ISO/TS 12913-2:2018, \isopl\ and \isoev\ are theoretically orthogonal axes, as observed by \cite{Axelsson2010}. Thus, the absence of significant differences in \isoev\ serves as a validation of the circumplex model and underscores the efficacy of the AMSS, which did not inadvertently impact \isoev.

The AMSS intervention demonstrated its restorative potential through a significant increase in the \textit{Fascination}, \textit{Being-Away}, and \textit{Compatibility} PRSS dimensions. Particularly noteworthy was the \SI{21.22}{\percent} increase in \fas, indicating the maskers' ability to captivate attention involuntarily \citep{Payne2018}, reinforcing the restorative effect of AMSS's informational masking mechanism \citep{Jeon2023}. Moreover, the significant shift of \textit{Being-Away} (\ba) from negative to positive suggests AMSS effectively transformed the traffic-exposed \rtgp\ soundscape from one associated with daily stressors to a source of respite \citep{Payne2018, Payne2013}.

Nevertheless, while the rise in \textit{Compatibility} (\com) due to AMSS was significant, its negative score fell short of expectations afforded by natural soundscapes such as waterfronts or vast green spaces \cite{Jeon2023}. The restorative limits of AMSS were also evident in both \textit{Extent-Coherence} and \textit{Extent-Scope} sub-dimensions. Despite a significant increase in natural sound dominance, perceived coherency (\ec) and expansiveness (\es) of the environment were unaffected, suggesting other factors, like visual impressions, may require adjustment. Notably, low \ec\ and \es\ scores are characteristic of urban environments \citep{Jeon2023}, consistent with observations in the \gfp. With significant correlations between \isopl\ and all PRSS dimensions except \es\ (Extended Data Table~\ref{tab:kendall_corrmat}), the positive link suggests the potential for the AMSS to enhance PRSS alongside \isopl, minimizing the need for separate models.

\subsection{Impact of AMSS through (psycho)acoustic metrics and their relation with perceptual factors} \label{sec:disc_amss_objective}

The disconnection between (psycho)acoustic parameters and restorative indicators (i.e. \fas, \ba) contrasts with \cite{Jeon2023}, where \laeq\ correlated negatively with both \fas\ and \ba, albeit with a brief \SI{3}{\minute} stimuli exposure time in \cite{Jeon2023}. This highlights the challenge of using objective metrics to assess the restorative impact of augmenting ``wanted'' sounds in noisy environments.

Considering that the \gfp\ had a mean \laeq\ about \dba{6} lower \chadded[comment=R1.3]{than \rtgp} (Section~\ref{sec:obj}), this suggests that AMSS augmentation \chadded[]{may correspond to a perceived \dba{6} noise reduction in terms of \isopl\ and PRSS dimensions}. However, unlike previous lab experiments \chadded[]{where \pln\ was significantly reduced with the addition of bird maskers \citep{Hong2020h,Hong2021b,Lam2023}, AMSS did not reduce the \pln\ at \rtgp. The increase in pleasantness without a corresponding reduction in \pln\ was also noted in a study by \citet{Hao2016}. It is worth noting that the \pln\ of traffic can be influenced by the visibility of the traffic source, with noise appearing louder when the source is not visible \citep{Hong2020b}. Additionally, the difference in \pln\ perception compared with earlier studies could be due to the longer \SI{10}{\minute} exposure duration in this study, as opposed to \SI{3}{\minute} in \citep{Hong2021b}, \SI{30}{\second} in \citep{Lam2023,Hao2016}, and \SI{10}{\second} in \citep{Hong2020h}. Hence, these findings suggest} that (psycho)acoustic parameters alone are unable to fully capture soundscape perception changes \chadded[]{and restorative potential.}

Limitations of objective parameters in predicting subjective responses to soundscape augmentation were highlighted in an indoor experiment \citep{Lam2023}, where perceived annoyance was more accurately predicted by \lceq\ and \isopl\ than by objective parameters alone. Similarly, while \nnf\ accurately predicted perceived traffic noise loudness indoors \citep{Lam2023}, this did not hold true in this outdoor study, highlighting the need for caution in direct comparisons due to limited data (\textit{sites} and \textit{conditions}).



\subsection{Limitations and opportunities}\label{sec:limitations}

The AI model in the AMSS used only acoustic data to determine the optimal masker-gain combinations \citep{Watcharasupat2022}. However, factors such as participant demographics and visual environment could influence perception \citep{Erfanian2021,Li2020}. \chadded[comment=R2.3]{While our linear mixed-effects model accounted for individual baseline differences in \isopl\ and PRSS dimensions by incorporating a random intercept for participants, the significant increase in these constructs reflects a consistent effect across participants on average, though not every individual necessarily experienced a significant increase. This variability is captured by the variance in \isopl\ and PRSS scores.}

\chadded[]{Although the current system was designed for public urban environments,} future AMSS versions could explore multimodal models, incorporating participant-linked information and real-time visual data for broader applicability in different contexts \cite{Karnapi2021,Ooi2023b}. \chadded[comment=R2.3]{Additionally, while demographic factors such as age and gender appeared less influential on system performance, education level, housing type, and noise sensitivity had a more significant impact on predictions, suggesting that future systems could benefit from incorporating these factors for more personalized experiences \citep{Ooi2023b}.} Since \isoev\ is orthogonal to \isopl, future models could \chadded[]{also} optimize changes in \isoev\ or a combination of both.

Drawing from an extensive survey and catalog of global soundscape interventions \citep{Moshona2022,Moshona2023}, the AMSS stands out as the only AI-based intervention specifically engineered to autonomously elevate \isopl\ levels. Notably, among AI models trained on comprehensive datasets adhering to ISO 12913 standards \citep{Hou2023,Hou2023a,Mitchell2021b}, the AMSS hosts the sole built-in prediction model capable of probabilistic modeling of \isopl. The cloud-based framework behind AMSS could potentially streamline soundscape interventions and monitoring on a large scale. It holds the key to cost-effective, large-scale perceptual mapping compared to traditional methods reliant on human responses \citep{Mitchell2023,Jiang2022}. This advancement could address challenges in widespread ISO 12913 standards adoption, particularly in predicting the socioeconomic impact of soundscapes \citep{Aletta2024,Jiang2022}.

%% file: body/05conclusion.tex
\section{Conclusion}

In conclusion, we described the implementation and validation of an AI-based soundscape augmentation system (the AMSS) deployed at a pavilion at which road traffic was the dominant noise source in the acoustic environment. Although the AMSS was designed only to select maskers for playback that maximized the \isopl\ of the deployment location, we found corresponding improvements in the rated overall quality, perceived restorativeness, appropriateness, and positive affect by the participants in the validation study. The \isopl\ of the deployment location was also found to have increased to a level similar to that of a different pavilion where road traffic was significantly less dominant, and where the objectively-measured SPL was signficantly lower. This was despite the fact that the AMSS caused a slight increase in objectively-measured SPL at the deployment location due to the playback of maskers via a four-speaker system. 

In addition, the AMSS requires no human input to run, thereby allowing for reductions in time and labor required to pick suitable maskers for augmentation as compared to traditional approaches involving expert guidance or post hoc analysis of study results. The physical hardware of the AMSS was also installed after the pavilions had been built, with minimal alterations to the surrounding environment and infrastructure. Therefore, there is great potential to further develop the AMSS and its corresponding soundscape augmentation approach for sustainable management of noise pollution, especially in built-up areas where physical modifications to the surroundings to manage noise may be impractical or unfeasible.

%% file: body/98endmatter.tex
\section*{Declaration of competing interest}

The authors declare that they have no known competing financial interests or personal relationships that could have appeared to influence the work reported in this paper.

\section*{Acknowledgements}

\db{This research is supported by the Singapore Ministry of National Development and the National Research Foundation, Prime Minister's Office under the Cities of Tomorrow Research Programme (Award No. COT-V4-2020-1). Any opinions, findings and conclusions or recommendations expressed in this material are those of the authors and do not reflect the view of National Research Foundation, Singapore, and Ministry of National Development, Singapore. We would also like to thank the People's Association for their support in the participant recruitment process, and the Pasir Ris-Punggol Town Council for their assistance in the deployment of our system.

Part of this work was done while K. N. Watcharasupat was supported by the AAUW International Fellowship from the American Association of University Women (AAUW), and separately by the IEEE Signal Processing Society Scholarship.}

\section*{Data Availability}

The data that support the findings of this study are openly available in NTU research data repository DR-NTU (Data) at \url{https://doi.org/10.21979/N9/NEH5TR}. The replication code used in this study is available on GitHub at the following repository: \url{https://doi.org/10.5281/zenodo.11141691}. The code includes all the necessary scripts, functions, and instructions to reproduce the results reported in the study. \chadded[comment=R1.1]{Additionally, due to intellectual property protection, we are unable to provide the replication code for the PPAP \isopl\ prediction model. However, a baseline CNN-based \isopl\ prediction model is available at the following repository: \url{https://github.com/ntudsp/araus-dataset-baseline-models}}

\printcredits

%% file: body/99appendix.tex
\input{commands/appendix-start}
\onecolumn
\section{Questionnaires}
\label{sec:questions}
\hspace{2em}
\input{tables/stimuliquestion.tex}
\clearpage
\input{tables/prss_table}
\clearpage
\input{tables/participantquestion.tex}
\clearpage

\section{Statistical results}\label{sec:statresults}

\input{tables/kstest}

\input{tables/alllstatsresults}

\input{tables/corrmatrix_obj}

%% file: commands/appendix-start.tex
\setcounter{section}{0}
\renewcommand{\thesection}{Appendix \Alph{section}}

\setcounter{table}{0}
\renewcommand{\thetable}{\Alph{section}.\arabic{table}}

\setcounter{figure}{0}
\renewcommand{\thefigure}{\Alph{section}.\arabic{figure}}

%% file: tables/stimuliquestion.tex
\begin{sffamily}
\begin{nolinenumbers}
\setcounter{table}{0}
\refstepcounter{table}\label{tab:stimuliquestion}

\small

\noindent\textbf{\color{scolor}Table \thetable}\par%

\noindent{Site evaluation questionnaire for the assessment of the soundscapes at the two study sites \gfp\ and \rtgp. Participants completed this questionnaire after a \SI{10}{\minute} listening period at each site.}

\begin{longtable}{%
    @{\extracolsep{\fill}}%
    >{\raggedright\arraybackslash}p{0.15\textwidth}%
    >{\raggedright\arraybackslash}p{0.2\textwidth}%
    >{\raggedright\arraybackslash}p{0.35\textwidth}%
    >{\raggedright\arraybackslash}p{0.2\textwidth}%
    @{}
}

    \toprule
    \textbf{Question Category}
    & \textbf{Instructions/Question}
    & \textbf{Specific Items}
    & \textbf{Rating Scale/Format}\\
    \midrule
\endhead

    \multicolumn{4}{r}{[Continued on next page]} 
\endfoot

    \bottomrule
    
\endlastfoot

    \multirow[t]{10}{=}{International Positive and Negative Affect Schedule Short Form (I-PANAS-SF)}
    & \multirow[t]{10}{=}{
    Indicate to what extent you feel this way in this moment.
    } 
    & Active
    & \multirow[t]{10}{=}{Very slightly or not at all--Extremely (5-point categorical)}\\ \cline{3-3}

    & & Attentive
    & \\ \cline{3-3} 

    & & Alert
    & \\ \cline{3-3}

    & & Determined
    & \\ \cline{3-3}

    & & Inspired
    & \\ \cline{3-3}

    & & Hostile
    & \\ \cline{3-3}

    & & Ashamed
    & \\ \cline{3-3}

    & & Upset
    & \\ \cline{3-3}

    & & Afraid
    & \\ \cline{3-3}

    & & Nervous
    & \\ 

\midrule

    \multirow[t]{3}{=}{Perceived Sound Source Dominance (DOM)}
    & \multirow[t]{3}{=}{To what extent do you presently hear the following types of sounds?}
    & Noise (e.g., traffic, construction, industry) 
    & \multirow[t]{3}{=}{Not at all--Dominates completely (5-point categorical scale)}\\ \cline{3-3}

    & & Sounds from human beings (e.g., conversation, laughter, children at play, footsteps) & \\ \cline{3-3}

    & & Natural sounds (e.g., singing birds, flowing water, wind in vegetation) & \\ 
\midrule
    \multirow[t]{8}{=}{Perceived Affective Quality (PAQ)}
    & \multirow[t]{8}{=}{For each of the 8 scales below, to what extent do you agree or disagree that the surrounding sound environment you heard is $\cdots$} 
    & Eventful
    & \multirow[t]{8}{=}{Strongly disagree--Strongly agree (5-point Categorical scale)}\\ \cline{3-3}

    & & Vibrant & \\ \cline{3-3}

    & & Pleasant & \\ \cline{3-3}

    & & Calm & \\ \cline{3-3}

    & & Uneventful & \\ \cline{3-3}

    & & Monotonous & \\ \cline{3-3}

    & & Annoying & \\ \cline{3-3}

    & & Chaotic & \\

\midrule
\pagebreak
    Overall Soundscape Quality (\osq)
    & \multicolumn{2}{>{\RaggedRight\arraybackslash}p{0.55\textwidth}}{Overall, how would you describe the present surrounding sound environment?}
    & Very good--Very bad (5-point Categorical scale) \\
\midrule
    Appropriateness (\appr)
    & \multicolumn{2}{>{\RaggedRight\arraybackslash}p{0.55\textwidth}}{Overall, to what extent is the present surrounding sound environment appropriate to the present place?}
    & Not at all--Perfectly (5-point Categorical scale) \\
\midrule
    Perceived Loudness (\pln)
    & \multicolumn{2}{>{\RaggedRight\arraybackslash}p{0.55\textwidth}}{How loud would you say the sound environment is?}
    & Not at all--Extremely (5-point Categorical scale) \\
\midrule
    \multirow[t]{4}{=}{Perceived Restorativeness Soundscape Scale (PRSS) -- Fascination (\fas)}
    & \multirow[t]{4}{=}{How much do you agree with the following statements?}
    & My curiosity is awoken by these sounds 
    & \multirow[t]{4}{=}{Not at all--Completely (7-point categorical scale)}\\ \cline{3-3}

    & & There are plenty of sounds for me to discover & \\ \cline{3-3}

    & & These sounds, I find fascinating & \\ \cline{3-3}

    & & My interest is really held by following what is going on with these sounds & \\ 
\midrule
    \multirow[t]{5}{=}{PRSS -- Being-away (\ba)} 
    & \multirow[t]{5}{=}{How much do you agree with the following statements?}
    
    & I get a break from my day-to-day routine from spending time with these sounds 
    
    & \multirow[t]{5}{=}{Not at all--Completely (7-point categorical scale)}\\ \cline{3-3}

    & & I find that I don't have to concentrate much when I'm surrounded by these sounds & \\ \cline{3-3}

    & & The sounds give me a chance to step back from things that demand my focus & \\ \cline{3-3}

    & & I feel free from work and/or responsibilities when I am with these sounds & \\ \cline{3-3}

    & & These sounds are a refuge for me from unwanted distractions 
    & \\ 
\midrule
\pagebreak
    \multirow[t]{3}{=}{PRSS -- Compatibility (\com)}
    & \multirow[t]{3}{=}{How much do you agree with the following statements?}
    
    & I rapidly adapt to these sounds 
    
    & \multirow[t]{3}{=}{Not at all--Completely (7-point categorical scale)}\\ \cline{3-3}

    & & While I am with these sounds, it is easy to do what I want & \\ \cline{3-3}

    & & The sounds fit well with my preferences & \\ 
    
\midrule
    \multirow[t]{3}{=}{PRSS -- Extent-Coherence (\ec)}
    & \multirow[t]{3}{=}{How much do you agree with the following statements?}
    & The existing sounds belong to this soundscape  
    & \multirow[t]{3}{=}{Not at all--Completely (7-point categorical scale)}\\ \cline{3-3}

    & & The sounds blend together to create a harmonious soundscape & \\ \cline{3-3}

    & & The sounds in this environment are well-organized, which makes it easy for me to hear the relationships between them & \\ 
 \midrule
 \multirow[t]{3}{=}{PRSS -- Extent-Scope (\es)}
    & \multirow[t]{3}{=}{How much do you agree with the following statements?}   
    & There are lots of different sounds to explore in this place 
    & \multirow[t]{3}{=}{Not at all--Completely (7-point categorical scale)}\\ \cline{3-3}
    & & The sounds make it feel like this place is vast & \\ \cline{3-3}
    & & These sounds have the quality to create a world of their own & \\
    

\end{longtable}
\end{nolinenumbers}
\end{sffamily}

%% file: tables/prss_table.tex

\begin{table}[]
\caption{Derivation of the Perceived Restorative Soundscape Scale (PRSS) items}
\label{tab:prss}
\begin{tabularx}{1\textwidth}{%
>{\raggedright\arraybackslash}X%
*{2}{>{\raggedright\arraybackslash}p{0.38\textwidth}}%
>{\raggedright\arraybackslash}p{4em}%
}
\toprule
\textbf{PRSS Dimensions} 
&\textbf{PRSS Items (specific framing in \citet{Payne2018})} 
&\textbf{PRSS Items (this study)} 
&\textbf{Remarks} \\ 

\midrule
  
Fascination 
&\multicolumn{2}{c}{My curiosity is awoken by these sounds} 
&- \\ \cmidrule(l){2-4} 
 
&\multicolumn{2}{c}{There are plenty of sounds for me to discover} 
&- \\ \cmidrule(l){2-4} 
&\multicolumn{2}{c}{These sounds, I find fascinating} 
&- \\ \cmidrule(l){2-4} 
&\multicolumn{2}{c}{My interest is really held by following what is going on with these sounds} 
&- \\ 
\midrule

Being-Away 
&\multicolumn{2}{c}{I get a break from my day-to-day routine from spending time with these sounds} 
&- \\ \cmidrule(l){2-4} 
&My concentration is demanded by these sounds 
&I find that I don't have to concentrate much when I'm surrounded by these sounds 
&rephrased \\ \cmidrule(l){2-4} 
&From these sounds, I experience few attentional demands 
&The sounds give me a chance to step back from things that demand my focus 
&rephrased \\ \cmidrule(l){2-4} 
&\multicolumn{2}{c}{I feel free from work and/or responsibilities when I am with these sounds} 
&- \\ \cmidrule(l){2-4} 
&I need to think of my obligations when I am with these sounds 
&- 
&\multirow{2}{*}{removed} \\ \cmidrule(lr){2-4}
&\multicolumn{2}{c}{These sounds are a refuge for me from unwanted distractions} 
& \\ 
\midrule

Compatibility 
&There is an accordance between these sounds and what I like to do 
&- 
&removed \\ \cmidrule(l){2-4} 

&\multicolumn{2}{c}{I rapidly adapt to these sounds} 
&- \\ \cmidrule(l){2-4} 
&\multicolumn{2}{c}{While I am with these sounds, it is easy to do what I want} 
&- \\ \cmidrule(l){2-4} 
&My personal inclinations fits with being with these sounds 
&The sounds fit well with my preferences 
&rephrased \\ 
\midrule

Extent (Coherence) 
&\multicolumn{2}{c}{The existing sounds belong to this soundscape} 
&- \\ \cmidrule(l){2-4} 
&The sounds fit together to form a coherent soundscape 
&The sounds blend together to create a harmonious soundscape 
&rephrased \\ \cmidrule(l){2-4} 
&These sounds are coherent 

&\multirow{3}{0.38\textwidth}{The sounds in this environment are well-organized, which makes it easy   for me to hear the relationships between them} 
&\multirow{3}{*}{combined} \\ \cmidrule(l){2-2}
&The sounds are clearly organized 
& & \\ \cmidrule(l){2-2}
&The physical arrangement of these sounds has a clear order 
& & \\ 
\midrule

Extent (Scope) 
&There are plenty of sounds to allow exploration in many directions 
&There are lots of different sounds to explore in this place 
&rephrased \\ \cmidrule(l){2-4} 

&The extent of these sounds seems limitless 
&\multirow{2}{*}{The sounds make it feel like this place is vast} 
&\multirow{2}{*}{combined} \\ \cmidrule(l){2-2}

&These sounds feel very spacious 
& & \\ \cmidrule(l){2-4} 
&These sounds have the quality of being a whole world to themselves &These sounds have the quality to create a world of their own 
&rephrased \\ 
\bottomrule
\end{tabularx}%
\end{table}

%% file: tables/participantquestion.tex
\begin{sffamily}
\begin{nolinenumbers}
\refstepcounter{table}\label{tab:participantquestion}

\small

\noindent\textbf{\color{scolor}Table \thetable}\par%

\noindent{Participant information questionnaire administered prior to the end of each session. Participants completed this questionnaire after the soundscape evaluations had been completed at both study sites \chreplaced[comment=R1.2 R2.4]{\rtgp}{RTGP} and \chreplaced[]{\gfp}{GFP}.}

\begin{longtable}{%
    @{\extracolsep{\fill}}%
    >{\raggedright\arraybackslash}p{0.15\textwidth}%
    >{\raggedright\arraybackslash}p{0.2\textwidth}%
    >{\raggedright\arraybackslash}p{0.35\textwidth}%
    >{\raggedright\arraybackslash}p{0.2\textwidth}%
    @{}
}

    \toprule
    \textbf{Question Category}
    & \textbf{Instructions/Questions}
    & \textbf{Specific Items}
    & \textbf{Rating Scale/Format}\\
    \midrule
\endhead

    \multicolumn{4}{r}{[Continued on next page]} 
\endfoot

    \bottomrule
    
\endlastfoot

    Gender
    & \multicolumn{2}{>{\RaggedRight\arraybackslash}p{0.55\textwidth}}{What is your gender?}
    & Male/\hspace{0mm}Female/\hspace{0mm}Non-\hspace{0mm}conforming/\hspace{0mm}Prefer not to say \\
\midrule
    Age
    & \multicolumn{2}{>{\RaggedRight\arraybackslash}p{0.55\textwidth}}{What is your age?}
    & Integer in [21,70] \\
\midrule
    Occupation
    & \multicolumn{2}{>{\RaggedRight\arraybackslash}p{0.55\textwidth}}{What is your occupational status?}
    & Employed/\hspace{0mm}Unemployed/\hspace{0mm}Retired/\hspace{0mm}Student/\hspace{0mm}Rather not say/\hspace{0mm}Other \\
\midrule
    \multirow[t]{3}{=}{Individual Noise Sensitivity (INS)}
    & \multirow[t]{3}{=}{Select the option that best represents your level of agreement with the statement.}
    & I wouldn't mind living on a noisy street if the apartment I had was nice.
    & \multirow[t]{3}{=}{Strongly disagree--Strongly agree (5-point categorical scale)} \\ \cline{3-3}

    & & I am more aware of noise than I used to be. & \\ \cline{3-3} 
    & & No one should mind much if someone turns up his stereo full blast once in a while. & \\ \cline{3-3}
    & & At movies, whispering and crinkling candy wrappers disturb me. & \\ \cline{3-3} 
    & & I am easily awakened by noise. & \\ 
    & & If it's noisy where I'm studying, I try to close the door or window or move someplace else. & \\ \cline{3-3} 
    & & I get annoyed when my neighbors are noisy. & \\ \cline{3-3} 
    & & I get used to most noises without much difficulty. & \\ \cline{3-3}
    & & How much would it matter to you if an apartment you were interested in renting was located across from a fire station? & \\ \cline{3-3} 
    & & Sometimes noises get on my nerves and get me irritated. & \multirow[t]{3}{=}{Strongly disagree--Strongly agree (5-point categorical scale)} \\ \cline{3-3} 
    & & Even music I normally like will bother me if I'm trying to concentrate. & \\ \cline{3-3} 
    & & It wouldn't bother me to hear the sounds of everyday living from neighbors (footsteps, running water, etc). & \\ \cline{3-3}
    & & When I want to be alone, it disturbs me to hear outside noises. & \\ \cline{3-3} 
    & & I'm good at concentrating no matter what is going on around me. & \\ \cline{3-3}
    & & In a library, I don't mind if people carry on a conversation if they do it quietly. & \\ \cline{3-3}
    & & There are often times when I want complete silence. & \\ \cline{3-3} 
    & & Motorcycles ought to be required to have bigger mufflers. & \\ \cline{3-3} 
    & & I find it hard to relax in a place that's noisy. & \\ \cline{3-3} 
    & & I get mad at people who make noise that keeps me from falling asleep or getting work done. & \\ \cline{3-3} 
    & & I wouldn't mind living in an apartment with thin walls. & \\ \cline{3-3}
    & & I am sensitive to noise. & \\ 
\midrule
    \multirow[t]{8}{=}{Baseline Noise Annoyance (BNA)}
    & \multirow[t]{8}{=}{In general, how much does noise from \underline{\hspace{1.5cm}} bother, disturb, or annoy you?}
    & Aircraft (military or civilian) & \multirow[t]{8}{=}{Not at all--Extremely (5-point Categorical scale)} \\ \cline{3-3}
    & & Road traffic & \\ \cline{3-3}
    & & MRT (trains) & \\ \cline{3-3}
    & & Children & \\ \cline{3-3}
    & & Other people & \\ \cline{3-3}
    & & Animals & \\ \cline{3-3}
    & & Construction worksites & \\ \cline{3-3}
    & & Construction (renovations) & \\ \cline{3-3}
    & & Any other noises & \\
\midrule
    \multirow[t]{18}{=}{Perceived Stress Scale (PSS)}
    & \multirow[t]{18}{=}{In the last month, how often have you...} & been upset because of something that happened unexpectedly? & \multirow[t]{18}{=}{Never--Very often (5-point categorical scale)} \\ \cline{3-3}
    & & felt that you were unable to control the important things in your life? & \\ \cline{3-3}
    & & felt nervous and ``stressed''? & \\ \cline{3-3}
    & & felt confident about your ability to handle your personal problems? & \\ \cline{3-3}
    & & felt that things were going your way? & \\ \cline{3-3}
    & & found that you could not cope with all the things that you had to do? & \\ \cline{3-3}
    & & been able to control irritations in your life? & \\ \cline{3-3}
    & & felt that you were on top of things? & \\ \cline{3-3}
    & & been angered because of things that were outside of your control? & \\ \cline{3-3}
    & & felt difficulties were piling up so high that you could not overcome them? & \\
\midrule
    \multirow[t]{5}{=}{WHO-Five Well-Being Index (WHO-5)}
    & \multirow[t]{5}{=}{For each of these statements, which is the closest to how you have been feeling over the last two weeks?}
    & I have felt cheerful and in good spirits. & \multirow[t]{5}{=}{At no time--All of the time (6-point Categorical scale)} \\ \cline{3-3}
    & & I have felt calm and relaxed. & \\ \cline{3-3}
    & & I have felt active and vigorous. & \\ \cline{3-3}
    & & I woke up feeling fresh and rested. & \\ \cline{3-3}
    & & My daily life has been filled with things that interest me. & \\ 










\end{longtable}
\end{nolinenumbers}
\end{sffamily}

%% file: tables/kstest.tex
\setlength{\LTpost}{0mm}
\setcounter{table}{1}

\begin{table}[H]

\centering

\begin{minipage}{1\linewidth}
\caption{Summary of exact two-sample Kolmogorov-Smirnov tests to examine effect of order (\gfp--\rtgp\ or \rtgp--\gfp) and group size (1 or $>1$) on each soundscape evaluation attribute (sound source dominance, overall quality, appropriateness, loudness, \isopl, \isoev, and PRSS dimensions) across each condition (\amss\ and \amb). All the $p$-values were adjusted for multiple comparisons within conditions with the Benjamini-Hochberg (BH) method.} \label{tab:kstest}
\end{minipage}

\scriptsize
\begin{tabularx}{1\linewidth}{
>{\raggedright\arraybackslash}p{5em}%
*{15}{>{\raggedleft\arraybackslash}X}}
\toprule
 
& \noi
& \nat 
& \hum
& \osq
& \appr 
& \pln
& \isopl 
& \isoev 
& \pa
& \na
& \fas
& \ba
& \com
& \ec
& \es \\
\midrule\addlinespace[2.5pt]
\multicolumn{16}{l}{Order} \\ 
\midrule\addlinespace[2.5pt]
\amb\ 
&  1.00 &  1.00 &  1.00 &  1.00 &  1.00 &  1.00 &  1.00 &  1.00 &  1.00 &  1.00 &  1.00 &  1.00 &  1.00 &  1.00 &  1.00 \\ 

\amss\ 
&  0.95 &  0.83 &  0.83 &  0.83 &  0.83 &  0.83 &  0.83 &  0.83 &  0.83 &  0.83 &  0.83 &  0.83 &  0.83 &  0.83 &  0.83 \\

\midrule\addlinespace[2.5pt]
\multicolumn{16}{l}{Group Size} \\ 
\midrule\addlinespace[2.5pt]
\amb\ 
&  0.98 &  0.98 &  0.98 &  0.98 &  0.96 &  0.96 &  0.98 &  0.96 &  0.98 &  0.96 &  0.98 &  0.96 &  0.96 &  0.98 &  0.96 \\

\amss\ 
&  0.75 &  0.75 &  0.94 &  0.75 &  0.75 &  0.94 &  0.75 &  0.94 &  0.75 &  0.94 &  0.80 &  0.80 &  0.75 &  0.75 &  0.80 \\

\bottomrule

\end{tabularx}
\end{table}

%% file: tables/alllstatsresults.tex
\setcounter{table}{0}
\footnotesize
\begin{xltabular}{1\textwidth}{X%
>{\raggedleft\arraybackslash}p{4cm}%
>{\raggedleft\arraybackslash}p{2cm}%
>{\raggedleft\arraybackslash}p{2cm}%
>{\raggedleft\arraybackslash}p{2cm}%
}
\caption{Summary of statistical tests for attributes in soundscape evaluation questionnaire (sound source dominance, overall quality, appropriateness, loudness, \isopl, \isoev, and PRSS dimensions) across site (\gfp\ and \rtgp), condition (\amss\ and \amb), and their interaction (\textit{site}:\textit{condition}). Test abbreviations and symbols for significance levels and effect sizes are defined in the footnote.} \label{tab:statresults}
\footnotesize
\\
\toprule
\textbf{Term} 
& \textbf{Test}\textsuperscript{\textit{1}} 
& \textbf{Estimate}\textsuperscript{\textit{2}}
& $p-$\textbf{value}\textsuperscript{\textit{3}} 
& \textbf{Effect Size}\textsuperscript{\textit{4}} \\
\endfirsthead

\multicolumn{4}{r@{}}{\small Continued from the previous page\ldots}\\

\toprule
\textbf{Term} 
& \textbf{Test}\textsuperscript{\textit{1}} 
& \textbf{Estimate}\textsuperscript{\textit{2}}
& $p-$\textbf{value}\textsuperscript{\textit{3}} 
& \textbf{Effect Size}\textsuperscript{\textit{4}} \\
\midrule
\endhead
\multicolumn{5}{r@{}}{\small Continues to the next page\ldots}\\
\endfoot
\endlastfoot

\midrule\addlinespace[2.5pt]
\multicolumn{5}{l}{\textbf{Sound source dominance -- Noise (\noi)}} \\ 

site & \rtrmanova & -15.1259 & \textbf{****0.0000} & (L)0.3182 \\ 
condition & \rtrmanova & 4.5451 &  0.1571 & (S)0.0145 \\ 
site:condition & \rtrmanova & -1.5148 &  0.5667 & 0.0000 \\ 

\midrule\addlinespace[2.5pt]
\multicolumn{5}{l}{\textbf{Sound source dominance -- Natural sounds (\nat)}} \\ 

site & \rtrmanova & 9.8815 & \textbf{***0.0004} & (L)0.1464 \\ 
condition & \rtrmanova & -9.7322 & \textbf{**0.0015} & (M)0.1175 \\ 
site:condition & \rtrmanova & 9.9857 & \textbf{***0.0003} & (L)0.1492 \\ 

{\hspace{2em}AMB - AMSS | GND} & Simple Contrasts for Condition & 0.5069 &  0.9513 & (S)0.0149 \\ 
{\hspace{2em}AMB - AMSS | ROOF} & Simple Contrasts for Condition & -39.4358 & \textbf{****0.0000} & -1.1574 \\ 
{\hspace{2em}GND - ROOF | AMB} & Simple Contrasts for Site & 39.7344 & \textbf{****0.0000} & (L)1.1661 \\ 
{\hspace{2em}GND - ROOF | AMSS} & Simple Contrasts for Site & -0.2083 &  0.9783 & -0.0061 \\ 

\midrule\addlinespace[2.5pt]
\multicolumn{5}{l}{\textbf{Sound source dominance -- Human sounds (\hum)}} \\ 

site & \rtrmanova & 26.1128 & \textbf{****0.0000} & (L)0.5180 \\ 
condition & \rtrmanova & 3.5269 &  0.1039 & (S)0.0121 \\ 
site:condition & \rtrmanova & -0.3316 &  0.8785 & 0.0000 \\ 

\midrule\addlinespace[2.5pt]
\multicolumn{5}{l}{\textbf{Positive Affect (\pa)}} \\

Residuals & Shapiro-Wilk normality test & - &  0.1731 &    - \\ 
site & \rmanova\ & 0.0256 &  0.6753 & 0.0000 \\ 
condition & \rmanova\ & -0.0712 &  0.1620 & (S)0.0139 \\ 
site:condition & \rmanova\ & -0.0003 & *0.0211 & (S)0.0403 \\ 
{\hspace{2em}AMB - AMSS | GND} & Simple Contrasts for Condition & -0.0024 &  0.9835 & -0.0050 \\ 
{\hspace{2em}AMB - AMSS | MP} & Simple Contrasts for Condition & -0.1431 &  0.2242 & -0.2963 \\ 
{\hspace{2em}AMB - AMSS | ROOF} & Simple Contrasts for Condition & -0.2819 & \textbf{*0.0179} & -0.5839 \\ 
{\hspace{2em}GND - MP | AMB} & Simple Contrasts for Site & 0.0323 &  0.8971 & (M)0.0669 \\ 
{\hspace{2em}GND - ROOF | AMB} & Simple Contrasts for Site & 0.1406 &  0.1369 & (L)0.2912 \\ 
{\hspace{2em}MP - ROOF | AMB} & Simple Contrasts for Site & 0.1083 &  0.2999 & (L)0.2243 \\ 
{\hspace{2em}GND - MP | AMSS} & Simple Contrasts for Site & -0.1083 &  0.2625 & -0.2243 \\ 
{\hspace{2em}GND - ROOF | AMSS} & Simple Contrasts for Site & -0.1389 &  0.1133 & -0.2876 \\ 
{\hspace{2em}MP - ROOF | AMSS} & Simple Contrasts for Site & -0.0306 &  0.8977 & -0.0633 \\ 

\midrule\addlinespace[2.5pt]

\pagebreak
\\ \multicolumn{5}{l}{\textbf{Negative Affect (\na)}} \\ 

Residuals & Shapiro-Wilk normality test & - & \textbf{****0.0000} &    - \\ 
site & \rtrmanova & -0.7847 &  0.3525 & 0.0006 \\ 
condition & \rtrmanova & 10.4994 & \textbf{*0.0253} & (S)0.0550 \\ 
site:condition & \rtrmanova & 3.9746 &  0.1665 & (S)0.0114 \\ 

\midrule\addlinespace[2.5pt]
\multicolumn{5}{l}{\textbf{Overall soundscape quality (\osq)}} \\ 
site & \rtrmanova & 7.7782 & \textbf{**0.0041} & (M)0.0965 \\ 
condition & \rtrmanova & -4.3164 &  0.2204 & 0.0073 \\ 
site:condition & \rtrmanova & 5.9796 & \textbf{*0.0271} & (S)0.0540 \\ 
{\hspace{2em}AMB - AMSS | GND} & Simple Contrasts for Condition & 3.3264 &  0.7087 & (M)0.0910 \\ 
{\hspace{2em}AMB - AMSS | ROOF} & Simple Contrasts for Condition & -20.5920 & *0.0221 & -0.5631 \\ 
{\hspace{2em}GND - ROOF | AMB} & Simple Contrasts for Site & 27.5156 & ***0.0009 & (L)0.7525 \\ 
{\hspace{2em}GND - ROOF | AMSS} & Simple Contrasts for Site & 3.5972 &  0.6297 & (M)0.0984 \\ 

\midrule\addlinespace[2.5pt]
\multicolumn{5}{l}{\textbf{Appropriateness (\appr)}} \\

site & \rtrmanova & 8.4062 & \textbf{**0.0024} & (M)0.1074 \\ 
condition & \rtrmanova & -10.8611 & \textbf{***0.0007} & (M)0.1327 \\ 
site:condition & \rtrmanova & 3.9062 &  0.1591 & (S)0.0142 \\ 

\midrule\addlinespace[2.5pt]
\multicolumn{5}{l}{\textbf{Perceived loudness (\pln)}} \\

site & \rtrmanova & -14.6107 & \textbf{****0.0000} & (L)0.3561 \\ 
condition & \rtrmanova & 1.9848 &  0.5667 & 0.0000 \\ 
site:condition & \rtrmanova & -5.3815 & *0.0221 & (S)0.0587 \\ 
{\hspace{2em}AMB - AMSS | GND} & Simple Contrasts for Condition & -6.7934 &  0.4189 & -0.1971 \\ 
{\hspace{2em}AMB - AMSS | ROOF} & Simple Contrasts for Condition & 14.7326 & 0.0812 & (L)0.4274 \\ 
{\hspace{2em}GND - ROOF | AMB} & Simple Contrasts for Site & -39.9844 & \textbf{****0.0000} & -1.1600 \\ 
{\hspace{2em}GND - ROOF | AMSS} & Simple Contrasts for Site & -18.4583 & \textbf{**0.0057} & -0.5355 \\ 

\midrule\addlinespace[2.5pt]

\pagebreak
\\ \multicolumn{5}{l}{\textbf{ISO Pleasantness (\isopl)}} \\

Residuals & Shapiro-Wilk normality test & - &  0.1229 &    - \\ 
site & \rmanova\ & 0.0966 & \textbf{**0.0011} & (M)0.1248 \\ 
condition & \rmanova\ & -0.0681 & \textbf{*0.0432} & (S)0.0434 \\ 
site:condition & \rmanova\ & 0.0781 & \textbf{**0.0082} & (M)0.0808 \\ 
{\hspace{2em}AMB - AMSS | GND} & Simple Contrasts for Condition & 0.0200 &  0.8241 & (S)0.0541 \\ 
{\hspace{2em}AMB - AMSS | ROOF} & Simple Contrasts for Condition & -0.2923 & \textbf{**0.0014} & -0.7926 \\ 
{\hspace{2em}GND - ROOF | AMB} & Simple Contrasts for Site & 0.3494 & \textbf{***0.0001} & (L)0.9473 \\ 
{\hspace{2em}GND - ROOF | AMSS} & Simple Contrasts for Site & 0.0371 &  0.6487 & (M)0.1006 \\ 

\midrule\addlinespace[2.5pt]
\multicolumn{5}{l}{\textbf{ISO Eventfulness (\isoev)}} \\ 

Residuals & Shapiro-Wilk normality test & - &  0.7790 & - \\ 
site & \rmanova\ & -0.0150 &  0.4576 & 0.0000 \\ 
condition & \rmanova\ & -0.0118 &  0.5795 & 0.0000 \\ 
site:condition & \rmanova\ & 0.0000 &  0.9990 & 0.0000 \\ 

\midrule\addlinespace[2.5pt]
\multicolumn{5}{l}{\textbf{Perceived Restorativeness Soundscape Scale: \textit{Fascination} (\fas)}} \\ 

Residuals & Shapiro-Wilk normality test & - &  0.8728 &  - \\ 
site & \rmanova\ & 0.0762 & \textbf{*0.0203} & (M)0.0606 \\ 
condition & \rmanova\ & -0.1257 & \textbf{**0.0034} & (M)0.1000 \\ 
site:condition & \rmanova\ & 0.0866 & \textbf{**0.0083} & (M)0.0806 \\ 
{\hspace{2em}AMB - AMSS | GND} & Simple Contrasts for Condition & -0.0781 &  0.4713 & -0.1755 \\ 
{\hspace{2em}AMB - AMSS | ROOF} & Simple Contrasts for Condition & -0.4245 & \textbf{***0.0001} & -0.9538 \\ 
{\hspace{2em}GND - ROOF | AMB} & Simple Contrasts for Site & 0.3255 & **0.0011 & (L)0.7314 \\ 
{\hspace{2em}GND - ROOF | AMSS} & Simple Contrasts for Site & -0.0208 &  0.8178 & -0.0468 \\ 

\midrule\addlinespace[2.5pt]
\multicolumn{5}{l}{\textbf{Perceived Restorativeness Soundscape Scale: \textit{Being-Away} (\ba)}} \\ 

Residuals & Shapiro-Wilk normality test & - &  0.7777 &    - \\ 
site & \rmanova\ & 0.0446 &  0.3081 & 0.0006 \\ 
condition & \rmanova\ & -0.1572 & \textbf{**0.0034} & (M)0.1005 \\ 
site:condition & \rmanova\ & 0.1025 & \textbf{*0.0193} & (M)0.0618 \\ 
{\hspace{2em}AMB - AMSS | GND} & Simple Contrasts for Condition & -0.1094 &  0.4309 & -0.1920 \\ 
{\hspace{2em}AMB - AMSS | ROOF} & Simple Contrasts for Condition & -0.5194 & \textbf{***0.0003} & -0.9116 \\ 
{\hspace{2em}GND - ROOF | AMB} & Simple Contrasts for Site & 0.2943 & \textbf{*0.0241} & (L)0.5165 \\ 
{\hspace{2em}GND - ROOF | AMSS} & Simple Contrasts for Site & -0.1157 &  0.3390 & -0.2031 \\

\midrule\addlinespace[2.5pt]
\pagebreak
\\
\multicolumn{5}{l}{\textbf{Perceived Restorativeness Soundscape Scale: \textit{Compatibility} (\com)}} \\ 

Residuals & Shapiro-Wilk normality test & - &  0.3328 &    - \\ 
site & \rmanova\ & 0.0813 & \textbf{***0.0009} & (M)0.1287 \\ 
condition & \rmanova\ & -0.0883 & \textbf{*0.0135} & (M)0.0698 \\ 
site:condition & \rmanova\ & 0.0489 & \textbf{*0.0456} & (S)0.0422 \\ 
{\hspace{2em}AMB - AMSS | GND} & Simple Contrasts for Condition & -0.0787 &  0.3652 & -0.2209 \\ 
{\hspace{2em}AMB - AMSS | ROOF} & Simple Contrasts for Condition & -0.2743 & \textbf{**0.0020} & -0.7697 \\ 
{\hspace{2em}GND - ROOF | AMB} & Simple Contrasts for Site & 0.2604 & \textbf{***0.0005} & (L)0.7308 \\ 
{\hspace{2em}GND - ROOF | AMSS} & Simple Contrasts for Site & 0.0648 &  0.3378 & (L)0.1819 \\ 

\midrule\addlinespace[2.5pt]
\multicolumn{5}{l}{\textbf{Perceived Restorativeness Soundscape Scale: \textit{Extent-Coherence} (\ec)}} \\ 

Residuals & Shapiro-Wilk normality test & - &  0.9051 &    - \\ 
site & \rmanova\ & 0.0772 & \textbf{**0.0015} & (M)0.1182 \\ 
condition & \rmanova\ & -0.1015 & \textbf{**0.0023} & (M)0.1089 \\ 
site:condition & \rmanova\ & 0.0309 &  0.2031 & 0.0090 \\ 

\midrule\addlinespace[2.5pt]
\multicolumn{5}{l}{\textbf{Perceived Restorativeness Soundscape Scale: \textit{Extent-Scope} (\es)}} \\ 

Residuals & Shapiro-Wilk normality test & - & 0.0581 &    - \\ 
site & \rmanova\ & 0.0742 & \textbf{**0.0010} & (M)0.1254 \\ 
condition & \rmanova\ & -0.0629 & \textbf{*0.0410} & (S)0.0446 \\ 
site:condition & \rmanova\ & 0.0326 &  0.1504 & (S)0.0155 \\ 

\bottomrule
\end{xltabular}

\begin{minipage}{0.95\textwidth}
\textsuperscript{\textit{1}} Two-way linear mixed effects repeated measures Type III ANOVA (\rmanova); Two-way linear mixed effects repeated measures Type III Rank-transformed ANOVA (\rtrmanova) \\
\textsuperscript{\textit{2}} Fixed effects estimate with reference to \gfp\ for \textit{site} and \amb\ for \textit{condition}. Contrast estimates are respective of the contrast term in the ``Term'' column.\\
\textsuperscript{\textit{3}} $\text{*}p<0.05$; $\text{**}p<0.01$; $\text{***}p<0.001$; $\text{****}p<0.0001$\\
\textsuperscript{\textit{4}} Partial Omega squared ($\omega^2_{p}$) for linear mixed effects and Cohen's $d$ for simple contrasts. (L) large effect $>0.14$ ; (M) medium effect $>0.06$; (S) small effect $>0.01$ \\
\end{minipage}

%% file: tables/corrmatrix_obj.tex
\setlength{\LTpost}{0mm}
\setcounter{table}{2}

\begin{table}[!ht]
\footnotesize
\centering
\caption{Kendall correlation matrix between all objective acoustic measures and perceptual attributes in the site evaluation questionnaire where the significance of each entry in the upper triangle is denoted with a Holm-adjusted $p$-value and each entry in the lower triangle is denoted with an unadjusted $p$-value. Asterisks indicate $\text{*}p<0.05$; $\text{**}p<0.01$; $\text{***}p<0.001$; $\text{****}p<0.0001$. The unit diagonal has been removed for clarity.}
\label{tab:kendall_corrmatobj}
\begin{tabularx}{1\textwidth}{
>{\raggedright\arraybackslash}p{4em}%
*{10}{>{\raggedleft\arraybackslash}X}%
}
\toprule
\multicolumn{1}{l}{} & \isopl & \osq & \pa & \pln & \fas & \ba & \com & \laeq & \lceq & \nnf \\ 
\midrule\addlinespace[2.5pt]
\isopl &  & ***0.64 & 0.31 & **-0.40 & **0.40 & ***0.56 & ***0.61 &  -0.22 &  -0.10 &  -0.18 \\ 
\osq & ***0.64 &  & 0.32 & ***-0.45 &  0.29 & ***0.49 & ***0.52 &  -0.19 &  -0.09 &  -0.16 \\ 
\pa & **0.31 & **0.32 &  &  -0.05 & **0.40 & ***0.44 & **0.40 &  0.00 &  -0.04 &  0.04 \\ 
\pln & ***-0.40 & ***-0.45 &  -0.05 &  &  -0.20 & -0.31 & ***-0.44 &  0.29 &  0.25 &  0.28 \\ 
\fas & ***0.40 & **0.29 & ***0.40 & -0.20 &  & ***0.59 & ***0.59 &  -0.09 &  -0.07 &  -0.11 \\ 
\ba & ***0.56 & ***0.49 & ***0.44 & **-0.31 & ***0.59 &  & ***0.71 &  -0.05 &  0.01 &  -0.05 \\ 
\com & ***0.61 & ***0.52 & ***0.40 & ***-0.44 & ***0.59 & ***0.71 &  &  -0.19 &  -0.12 &  -0.18 \\ 
\laeq & *-0.22 & .-0.19 &  0.00 & **0.29 &  -0.09 &  -0.05 & -0.19 &  & ***0.59 & ***0.68 \\ 
\lceq &  -0.10 &  -0.09 &  -0.04 & *0.25 &  -0.07 &  0.01 &  -0.12 & ***0.59 &  & ***0.47 \\ 
\nnf &  -0.18 &  -0.16 &  0.04 & **0.28 &  -0.11 &  -0.05 & -0.18 & ***0.68 & ***0.47 &  \\ 
\bottomrule
\end{tabularx}
\end{table}